
\magnification=\magstep1
\baselineskip=12pt
\centerline{{\bf Exact Black Hole and Cosmological Solutions in a}}
\centerline{{\bf Two-Dimensional Dilaton-Spectator Theory of Gravity}}

\vskip 2.5 true cm

\centerline{{\bf K.C.K. Chan(1) and R.B. Mann(1,2)}}

\vskip 2.5 true cm

\centerline{(1) {\bf Department of Physics}}
\centerline{{\bf University of Waterloo, Waterloo, Ontario, Canada N2L 3G1}}

\vskip 0.5 true cm

\centerline{(2) {\bf Department of Applied Mathematics and Theoretical
Physics}}
\centerline{{\bf University of Cambridge, Silver St., Cambridge, England CB3
9EW}}

\vskip 2.5 true cm

\centerline{{\bf Abstract}}

Exact black hole and cosmological solutions are obtained for
a special two-dimensional dilaton-spectator ($\phi-\psi$) theory of gravity.
We show how in this context any desired spacetime behaviour
can be determined by an appropriate choice of a dilaton potential
function $V(\phi)$ and
a ``coupling function'' $l(\phi)$ in the action.
We illustrate several black hole solutions as examples.
In particular, asymptotically flat double- and multiple-
horizon black hole solutions are obtained. One solution bears an interesting
resemblance to the $2D$ string-theoretic black hole and contains the same
thermodynamic properties; another  resembles the $4D$
Reissner-Nordstrom solution. We find two characteristic features
of all the black hole solutions. First the coupling constants in $l(\phi)$
must be set equal to constants of integration (typically the mass). Second,
the spectator field $\psi$ and its derivative $\psi^{'}$ both diverge at
any event horizon. A test particle with ``spectator charge" ({\it
i.e.}  one coupled either to $\psi$ or $\psi^{'}$), will therefore
encounter an infinite tidal force at the horizon or an
``infinite potential barrier'' located outside the horizon respectively.
We also compute the
Hawking temperature and entropy for our solutions. In $2D$ $FRW$ cosmology, two
non-singular solutions which resemble two exact solutions in $4D$
string-motivated
cosmology are obtained. In addition, we construct a singular model
which describes the $4D$ standard non-inflationary big bang cosmology
($big-bang\rightarrow radiation\rightarrow dust$). Motivated by the
similarities
between $2D$ and $4D$ gravitational field equations
in $FRW$ cosmology, we briefly discuss a special
$4D$ dilaton-spectator action constructed from the bosonic part
of the low energy heterotic string action and get an exact solution
which contains dust and radiation behaviour.

\vskip 0.6 true cm

\noindent{\bf{1 Introduction}}

\vskip 0.2 true cm

In two spacetime dimensions, the Einstein action is a topological
invariant and so has no dynamical content.
Over the years many attempts have been made to formulate
a non-trivial theory of two-dimensional gravity (see, for
example, [1]). Such a study is not only pedagogically rewarding
for classical relativists [2], but can also yield insights into the behaviour
of semi-classical and quantum gravity [3]. This is primarily because
these theories are mathematically
simpler than four-dimensional general relativity ($GR$) yet retain
much of the conceptual complexity of $(3+1)$ dimensional spacetime
physics.

Recently two such theories, that of [3], referred to as the ``$R=T$''
theory, and the string-inspired theory of [4], have attracted some
interest, due to primarily to the fact that their field equations
admit exact black hole and cosmological solutions, making them an
interesting arena for the study of gravitational effects in both
classical and quantum regimes.
The former theory ($R=T$) retains the Einsteinian
``$curvature=matter$'' notion; indeed this theory
can be constructed by taking the
$D\rightarrow 2$ limit of $D$-dimensional $GR$ [5], or
by taking the limit $\omega\rightarrow\infty$ for the Brans-Dicke
parameter $\omega$ in the $2D$ Brans-Dicke theory [6]. In this sense
the $R=T$ theory can be viewed as a two-dimensional version of
$GR$. This viewpoint is further supported
by results that indicate the theory has a number of features which are
closely analogous to four-dimensional $GR$. These include a
well-defined Newtonian limit, post-Newtonian expansion
and gravitational collapse to a black hole [7].
Semi-classical calculations indicate
that the black hole radiates spinor [8] and vector particles [9].
Its classical cosmological properties are also similar
to the four-dimensional counter-parts [7,10].

In this paper, we search for new exact black hole and cosmological
solutions in the above-mentioned $R=T$ theory.
We start with the following two-dimensional generalized ``dilaton-
spectator'' action:
$$
S=S_G+S_M, \eqno(1a)
$$
where
$$
S_G=\int d^2x{\sqrt{-g}}(\tilde{\psi}R+{1\over 2}(\nabla\tilde\psi)^2),
\eqno(1b)
$$
$$
S_M=\int
d^2x{\sqrt{-g}}(a{\phi}R+H(\phi)(\nabla\phi)^2+h(\phi)F^2+l(\phi)(\nabla\psi)^2+V(\phi,\psi)). \eqno(1c)
$$
$S_G$ is the gravitational action and $S_M$ is the matter action. The latter by
definition
is independent of the auxiliary field $\tilde\psi$ which
has no effect on the classical evolution of the
gravity/matter system [2,3]. In $S_M$, $\phi$ has couplings analogous
to the usual dilaton field in other
2D theories; we shall therefore refer to it as the dilaton
and to  $\psi$ as the spectator field. $V(\phi,\psi)$ is the potential function
for $\phi$
and $\psi$. $F^2{\equiv}F^{\mu\nu}F_{\mu\nu}$ is the Maxwell contribution.

The action (1) contains several special cases. First of all,
when $S_G=H(\phi)=l(\phi)=h(\phi)=0$, $a=1$ and $V(\phi,\psi)=\phi\Lambda$,
one has the Jackiw-Teitelboim theory [11]. Second,
if $S_G=H(\phi)=l(\phi)=0$, $a=1$ and $V(\phi,\psi)=V(\phi)$,
one gets the kind of two-dimensional action which admits several
black hole solutions dimensionally reduced from the
three-dimensional ``BTZ'' charged and spinning black hole solution [12].
Third, when $S_G=H(\phi)=0$, $a=1$, one obtains the kind of
action considered by
Lechtenfeld and Nappi in the context of no hair theorem of black holes [13].
Fourth, just setting $S_G=0$ yields the action considered by Elizalde and
Odintsov on the discussion of one-loop renormalization and charged black
hole solutions [14]. In particular, if $a=1$, $h(\phi)=0$,
$H(\phi)={1\over\phi}$, $l(\phi)=-{2\over 3}\phi$ and $V(\phi,\psi)=
\lambda_1^2 e^{-2\phi}-\lambda_2^2e^{-{2b\over 3}\psi}$, then
one gets the action considered in [15] in a study of $2D$ black hole
radiation.

The aforementioned cases all
have $S_G=0$. Now if $S_G\neq 0$, we get the ``$R=T$'' theory.
For this ``R=T'' theory, when $a=1$, $H(\phi)=constant=2b$,
$h(\phi)=l(\phi)=0$ and $V(\phi,\psi)={\Lambda}e^{-2a\phi}$, (1) reduces
to the action of a Liouville field coupled to 2D gravity, where exact black
hole
solutions have been found [16]. In this paper, we will consider the ``$R=T$''
theory
({\it i.e.} $S_G \neq 0$) with matter action $S_M$ such that $a=0$,
$H(\phi)=2b$, and
$V(\phi,\psi)=V(\phi)$. Hence we consider
$$
S=S_G+\int
d^2x{\sqrt{-g}}(2b(\nabla\phi)^2+V(\phi)+l(\phi)(\nabla\psi)^2+h(\phi)F^2).
\eqno(2)
$$
Mathematically, $V(\phi)$ is a zero-form field, $\partial_\mu\psi$
is a one-form field (with dilaton coupling function $l(\phi)$)
and $F_{\mu\nu}$ is a two-form field (with dilaton coupling function
$h(\phi)$). Thus $S_M$ in (2) is of particular interest as it describes
a 2D dilaton theory of gravity which (i) has couplings of the dilaton to all
possible
$n$-form matter fields in 2D  and (ii) has its gravitational couplings manifest
via the  2D ``$curvature=matter$'' notion described earlier.

It is worthwhile pointing out that four-dimensional two-scalar field
gravitational theories are presently attracting much attention.  In
string gravity, in addition to the usual dilaton, a second scalar (called
the modulus field) describes the radius of the compactified space [17].
Furthermore, a class of multiple fields scalar-tensor theories of gravity and
cosmology has been studied in [18].
As pointed out by the authors in [13], the restriction to a single scalar
is merely for simplicity. For a more general situation, one should
consider a second scalar interacting with the usual dilaton.

We will show that the equations of motion of (2) in the static and
spatially homogeneous cases are very easy to solve if one adopts
a method which is the ``inverse'' of the usual
approach. Usually one specifies the form of $l(\phi)$, $h(\phi)$,
and $V(\phi)$ in the equations of motion
and then solves for the metric, $\phi$ and $\psi$;
that is, one first specifies the matter content,
then solves the field equations to determine the behaviour of the fields of
interest. In our approach we will do the
opposite: we show that if $l(\phi)$ is non-vanishing, then one can
first specify the metric and $\phi$ then solve
for $l(\phi)$, $h(\phi)$, $V(\phi)$ and $\psi$.
Indeed, the equations of motion permit almost any desired behaviour
for the metric provided  $l(\phi)$,
$h(\phi)$, $V(\phi)$ and $\psi$ are appropriately chosen.
The functions we obtain are {\sl{ad hoc}}
in the sense that they are derived from desired behaviour of the metric,
rather than from a two-dimensional field theory model.
However, this stance is not unusual in other studies. For example,
Ellis and Madsen adopted this approach in their
studies of four-dimensional exact scalar field cosmology [19].
Trodden, Mukhanov and Brandenberger obtained a non-singular
two-dimensional black hole solution in (1) for $S_G=0$,
$a=1$, $H(\phi)=l(\phi)=h(\phi)=0$ and $V(\phi,\psi)=V(\phi)$ [20].
They first specified the desired non-singular black hole metric and
then solved for $V(\phi)$.

Although such an approach for the action (2) offers a rather
straightforward way to obtain exact solutions, it has two
(perhaps unattractive)
characteristic features. First, for any black hole solution,
the spectator field $\psi$ and the term $l(\phi)(\nabla\psi)^2$
in (2) both diverge at the event horizon. We will show in detail
that if a test particle couples to $\psi$ in its equation of motion
({\it i.e.} has ``spectator charge'')
it will encounter an infinite tidal force at the horizon within a finite
proper time. Alternatively, if it
couples to $l(\phi)(\nabla\psi)^2$,
there exists an infinite potential barrier such that the test
particle starting its journey from infinity towards the event horizon
will be bounced back toward infinity
before reaching the horizon. For a neutral particle
(travelling along a geodesic), it will encounter no infinite tidal force
(except at the center) or infinite potential barrier. Second, we find
that no non-trivial exact solutions can be found unless (at least) one of the
coupling constants in the coupling function $l(\phi)$ functionally
determines (at least one)
an integration constant; typically this is the mass parameter of the black
hole solution. A similar situation occurs in
[21], in which a model of a two-dimensional
universe has a cosmological constant in a $2D$ gravitational action
appearing as a constant of integration in order to
get the desired behaviour of the universe;
namely, as the radius of the universe gets larger, the cosmological
constant gets smaller. An analogous $4D$ situation
was considered in [22] for
a deSitter (or anti-deSitter) black hole in which the cosmological constant
was (for whatever reason) set equal a function of the mass (and
charge, when relevant) -- these types of
solutions ({\sl{e.g.}} the extremal Schwarzschild-de Sitter
solution) form a set of measure zero in the space of all possible solutions.
In a more general approach, the $4D$ Einstein action may even be considered to
have
time-dependent coupling contants,
namely the gravitational and cosmological constants [23].

The action is, of course, the primary quantity in any physical theory.  The
black hole
solutions we obtain are therefore valid solutions to the field equations of the
actions we
consider only for black holes whose specific masses are
determined by the parameters of those
actions. The existence of other potentially interesting solutions associated
with these actions
remains an open question.  Hence the approach
we are using may be regarded as a method to suggest two dimensional actions
that are
of potential interest in 1+1 dimensions.
If we restrict ourselves to discussions of
a neutral test particle and adopt the aforementioned viewpoint regarding the
black hole solutions we obtain, then our solutions are physically interesting
since they admit double and multiple event horizons as examples, as well
as a couple of interesting non-singular cosmological solutions.

Our paper is organized as follows. In section 2 the equations of motion
associated with the action (2) are derived.
We discuss their properties and see how
to implement the desired behaviour of a given metric. In section 3
several interesting examples of black hole solutions
which are asymptotically flat with double or multiple
horizons are illustrated, and their quasi-local energy and mass are
also calculated. In addition, we construct a special
black hole solution which
has
 a causal structure
analogous to the Reissner-Nordstrom spacetime except that there are
no singularities. Section 4 is devoted to a discussion of
the motion of a test particle coupling to $\psi$ or $l(\phi)(\nabla\psi)^2$.
It is shown that the particle either encounters an infinite tidal force at the
event horizon or is bounced back toward infinity before reaching it.
The Hawking temperature and entropy for some of the black hole
solutions are computed in section 5. One solution in particular resembles
the $2D$ string black hole metric with the same thermodynamic
behaviour. A couple of
interesting non-singular $FRW$ cosmological models are extracted in section 6.
A singular universe which is analogous to the $4D$ standard big bang model
is also constructed. Motivated by the similarities in
field equations between two and four-dimensional $FRW$ cosmology,
we also briefly discuss a two-scalar field cosmology obtained from
$4D$ string theory. A dust/radiation solution is derived.
We summarize our work in the final section.
We set the $2D$ gravitational coupling
constant to be $1$ ({\sl{i.e.}} mass has a dimension of inverse
length) and the signature of the metric be $(-+)$.

\vskip 0.8 true cm

\noindent{\bf{2 Field Equations}}

\vskip 0.2 true cm

Varying (2) with respect to the
auxiliary, metric, dilaton, spectator and Maxwell fields yields
$$
{\nabla}^2\tilde\psi-R=0, \eqno(3)
$$
$$
{1\over 2}\Bigl(\nabla_\mu\tilde\psi\nabla_\nu\tilde\psi
-{1\over 2}g_{\mu\nu}(\nabla\tilde\psi)^2\Bigr)
+g_{\mu\nu}\nabla^2\tilde\psi-\nabla_\mu\nabla_\nu\tilde\psi
=T_{\mu\nu}, \eqno(4)
$$
$$
-4b\nabla^2\phi+{dl\over d\phi}(\nabla\psi)^2+{dh\over d\phi}F^2
+{dV\over d\phi}=0, \eqno(5)
$$
$$
\nabla^{\mu}(l(\phi)\nabla_{\mu}\psi)=0, \eqno(6)
$$
and
$$
\nabla^{\mu}(h(\phi)F_{\mu\nu})=0, \eqno(7)
$$
where
$$
T_{\mu\nu}={1\over 2}g_{\mu\nu}V(\phi)
-2h(\phi)\Bigl(F_{\mu\tau}{F_{\nu}}^{{}\tau}-{1\over 4}g_{\mu\nu}F^2\Bigr)
-2b\Bigl(\nabla_{\mu}\phi\nabla_{\nu}\phi-{1\over
2}g_{\mu\nu}(\nabla\phi)^2\Bigr)
$$
$$
-l(\phi)\Bigl(\nabla_\mu\psi\nabla_\nu\psi-{1\over
2}g_{\mu\nu}(\nabla\psi)^2\Bigr). \eqno(8)
$$
The above form a complete set of eight differential equations.
It is easy to see that $T_{\mu\nu}$ is conserved by taking the divergence of
(4);
along with the obvious gauge invariance of the Maxwell field, this shows that
there
are really only five independent field equations in (3)--(7).
Hence given $V(\phi)$, $l(\phi)$ and $h(\phi)$, then these differential
equations may be solved for the five unknown functions  $\phi$,
${\tilde{\psi}}$,
$\psi$, $F_{\mu\nu}$ and $g_{\mu\nu}$ where the latter two fields have
respectively
one independent function each due to gauge and co-ordinate invariance.

Insertion of the trace of (4) into (3) yields
$$
R=T \quad . \eqno(9)
$$
The evolution of $\tilde\psi$ is then determined only
by the traceless part of (4). Since $R$ completely determines the
metric field in two dimensions [1] and $T$ is independent of $\tilde\psi$,
we see that $\tilde\psi$ has no effect on the evolution of the gravity/matter
system
(although the converse is not true).

In two dimensions, a symmetric metric only has one degree of freedom [1].
Thus we can write the static metric in the form
$$
ds^2=-\lambda(x)dt^2+{dx^2\over\lambda(x)}. \eqno(10)
$$
Taking the form (10), the gravity-matter equation (9) reduces to
$$
\lambda^{''}=-V-{2q^2\over h}  \eqno(11a)
$$
where the $'$ denotes an ordinary derivative with respect
to $x$ and $q$ is the electric charge associated with the Maxwell field
(see (11d)).
One sees that the metric directly depends on the potential
and Maxwell contribution, not on the spatial derivative of the dilaton
or spectator. The spectator equation (7) becomes
$$
\psi^{'}={C_0\over\lambda l}, \eqno(11b)
$$
where $C_0$ is a constant of integration. The dilaton equation (5) is
$$
-4b(\lambda\phi^{'})^{'}+{C_0^2\over\lambda l^2}{dl\over d\phi}
+{d\over d\phi}\Bigl({2q^2\over h}+V\Bigr)=0. \eqno(11c)
$$
The Maxwell two-form is given by
$$
F_{\mu\nu}={q\over h}\epsilon_{\mu\nu}, \eqno(11d)
$$
where $q^2$ is the square of magnitude of an electric charge and
$\epsilon_{\mu\nu}$ is the volume two-form ($\epsilon^2=-2$).

We are now left with three differential
equations (11a)-(11c) from which one can solve
simultaneously for $\psi(x)$, $\phi(x)$
and $\lambda(x)$ for a given $V(\phi)$, $l(\phi)$ and $h(\phi)$.
Defining
$$
\tilde{V}=V+{2q^2\over h}, \eqno(12)
$$
(11a) and (11c) respectively become
$$
\lambda^{''}=-\tilde{V}, \eqno(13)
$$
and
$$
-4b(\lambda\phi^{'})^{'}+{C_0^2\over\lambda l^2}{dl\over d\phi}
+{d\tilde{V}\over d\phi}=0. \eqno(14)
$$
The definition in (12) is possible since the Maxwell
and the dilaton potential contributions are mathematically
equivalent in two dimensions.

{}From now on, we will drop the tilde sign in (13) and (14).
(3) always has the solution
$$
{\tilde\psi}^{'}=-{\lambda^{'}+C\over \lambda}. \eqno(15)
$$
$C$ is an integration constant which will
later be shown to be related to the mass parameter
in our static solutions. If $l(\phi)$ and $V(\phi)$ are specified, one has
to solve simultaneously three independent equations, (11b), (13) and (14)
for $\psi(x)$, $\lambda(x)$ and $\phi(x)$.

Differentiating both sides of (13) with respect to $x$,
multiplying by  $\lambda(x)$, and performing an integration
(with respect to $x$ again) yields
$$
\int\lambda dV=-\lambda\lambda^{''}+{1\over 2}\lambda^{'2}. \eqno(16)
$$
Provided $\phi^{'}\neq 0$, (14) can always be transformed to
$$
-4b(\lambda\phi^{'})^{'}+{l^{'}\over l^2}{C_o^2\over\lambda\phi^{'}}
+{V^{'}\over\phi^{'}}=0. \eqno(17)
$$
Integrating (17) yields (using (16))
$$
{C_0^2\over l}=-2b(\lambda\phi^{'})^2-\lambda\lambda^{''}
+{1\over 2}\lambda^{'2}+C_1, \eqno(18)
$$
where $C_1$ is a constant of integration.
In this paper, $C$ and $C_i$ ($i=0,1,2,...$) denote integration
constants in all the black hole and cosmological solutions.
Equations (11b), (13) and (18) now form the complete set of differential
equations to be solved. It is clear
that if an invertible $\phi=\phi(x)$
($x=x(\phi)$) and a metric field $\lambda(x)$ are
both specified {\it a-priori}, then (18) trivially yields $l(\phi)$.
Equation (13) then implies that $V(\phi)$ can be determined from the given
$\phi(x)$ and $\lambda(x)$. Here we take advantage of this
arbitrariness to determine $V(\phi)$ and $l(\phi)$.
In this sense we are determining which $V(\phi)$ and $\l(\phi)$
are required to yield a static metric in the $R=T$ theory of specified
form. One can see from (18)
that $C_0$ and $C_1$ are also coupling constants
in $l(\phi)$ for a given $\lambda(x)$ and $\phi(x)$.
Hence in this approach the solutions we obtain are valid only when
their constants of integration functionally depend upon the
coupling constants $C_0$ and $C_1$
in $l(\phi)$. In fact, $C_0$ is just a length scale with dimensions of inverse
length and (if nonzero) has no physical significiance.
It can be scaled to $1$ by adjusting
the units. However, it will be later shown that $C_1$
and $C$ are related to the mass parameter of a static black hole
solution. Therefore (18) indicates that each $l(\phi)$ we obtain yields
a special class of black hole solutions whose specific mass parameters
are determined by one of the coupling constants.
Henceforth, unless otherwise stated,
$C$ and $C_i$ ($i=0,1,2...$) will generally denote coupling constants in
action (2) that determine the integration constants for the
solutions we obtain via the above procedure; the rest of the constants
are just coupling constants with no such functional dependence.

Although this procedure seems somewhat similar to prescribing a
$4D$ metric, evaluating the resultant $G_{\mu\nu}$
and then determining the stress-energy
by setting it equal to this $G_{\mu\nu}$ via Einstein's equations ({\it i.e.}
one finds the matter required to give a desired geometric state), there
is a subtle distinction in that there are always appropriate dilaton potentials
which can be chosen to satisfy the field equations under the above constraints.
The solutions we obtain are those for which the constants of integration
functionally depend upon
the coupling constants in these potentials -- however other solutions to
the field equations may exist for which no such dependence exists.
Once a particular $l(\phi)$ and $V(\phi)$ have been determined,
one can then regard the action containing these functions as
a separate field theory in its own right and explore the solution
space of its field equations (one point of which must
contain the solution which originally
led to this choice of $l(\phi)$ and $V(\phi)$).

If $C_0$ vanishes, one can see in (18) that
it is generally difficult (if not impossible)
to integrate $\phi=\phi(x)$ and get its inverse if a
desired $\lambda(x)$ is given. Finally $\psi(x)$ is determined
by integrating (11b) or
$$
C_0\psi^{'}=-2b\lambda\phi^{'2}-\lambda^{''}+{1\over
2}{\lambda^{'2}\over\lambda}
+{C_1\over\lambda}, \eqno(19)
$$
where (18) has been used. Generally speaking, there exists
no general method for integrating (19) exactly. However, we see from the
action (2) and its equations of motion that none of them
depend explicitly or implicitly on $\psi(x)$ due to the fact
that we chose the potential $V$ to depend only on $\phi$. This choice
greatly simplifies our calculations and makes the exact integration
of
 $\psi(x)$
unnecessary (although in the cases discussed below
we find that $\psi^{'}(x)$ can be integrated exactly).
However, the price paid for this choice is that both
$\psi(x)$ and $\psi^{'}(x)$ diverge at the spatial point(s) where
$\lambda(x)=0$,
that is, at the event (or cosmological) horizons.
As a result, an infinite tidal force (for a test particle coupled to $\psi(x)$)
is present at the event horizon or an infinite potential barrier
(if it couples to $\psi^{'}$) exists outside the horizon.

Before proceeding further we note the following.
Besides the approach of obtaining exact solutions we mentioned
in the last paragraph, one can also adopt an alternate approach: given a
$V(\phi)$ and an invertible $\phi=\phi(x)$, (13) yields $\lambda(x)$.
Now substituting the $\lambda(x)$ into (18) yields $l(\phi)$. This approach
is less straightforward since $\lambda^{''}(x)$ is not always integrable
for a desired form of the potential. The third approach, (typically the
most difficult), is the usual approach: given $l(\phi)$ and $V(\phi)$, one
solves for $\lambda(x)$ and $\phi(x)$. Even in the present
simple two dimensional context, the system of differential
equations (13) and (18)
are typically difficult to solve in this third approach.
In this paper we adopt the first approach only.

\vskip 0.6 true cm

\noindent{\bf{3 Black Hole Solutions}}

\vskip 0.2 true cm

In this section, we first specify the forms of
$\lambda(x)$ and $\phi(x)$ and then solve for $l(\phi)$
and $V(\phi)$. To illustrate the procedure, we give some
special and interesting examples of black hole solutions.
We will not attempt to show all possible examples.
We note that four-dimensional $GR$ admits a static charged
black hole solution which is asymptotically flat and has a double-horizon
spacetime structure. In two-dimensional string-inspired theories of gravity,
asymptotically flat charged black hole solutions are even more interesting:
they have double as well as multiple horizons [24]. In $R=T$
theory, the situation is very different.
When $V=\psi=\phi=0$ and $h(\phi)=h_o$ in action (2), we have pure gravity
with a Maxwell contribution.
(11d) indicates that $F_{\mu\nu}$ is just a constant two form. The
photon field ${\tilde{A}}_{\mu}$
($F_{\mu\nu}=\partial_{[\mu}{\tilde{A}}_{\nu]}$)
propagates no physical degrees of freedom but the solution of its field
equation
allows an arbitrary constant $q^2$ in the gravitational field equation (11a).
Thus the Maxwell contribution (without $\phi$ and $\psi$) is equivalent
to a constant ${2q^2\over h_o}$ which appears as a constant of integration
in the equations of motion of action (2). Now the metric is given by [25]
$$
\lambda=-{q^2\over h_o}x^2+2mx\pm 1. \eqno(20)
$$
$m$ is the mass parameter. Although (20) admits a double-horizon spacetime,
it is not asymptotically flat. Indeed, $q^2$ plays the role as a $2D$
cosmological constant,
with the sign of $h_o$ determining whether the spacetime is asymptotically de
Sitter or
anti-de Sitter.
When a dilaton with a potential is present, no exact asymptotically flat
double-horizon black hole solutions are found either [26].
In order to see this, we use (18). We consider three types
of the simplest metric forms (i) $\lambda=A_1+{A_2\over x}+A_3x$,
(ii) $\lambda=A_1+{A_2\over x}+{A_3\over x^2}$, and
(iii) $\lambda=A_1+A_2e^{-kx}+A_3e^{-2kx}$. $A_j (j=1,2,3)$ are constants.
(i)-(iii) are asymptotically flat and admit double horizons for properly
chosen signs and magnitude among $A_j$. Note that in (i) the criterion for
asymptotically flatness is slightly more general than in higher dimensions.
One need only require that $\lambda\rightarrow A_1+A_3x$ for large
$\vert{x}\vert$ since in this case the metric will become asymptotically
like a Rindler space time; a Rindler transformation may then be applied
locally to rewrite the metric in Minkowski form [26].
Now putting (i)-(iii) into (18) when $C_0$ is vanishing,
it is not hard to check that $\phi^{'}(x)$ cannot be integrated exactly
to close forms and therefore no inverse $x=x(\phi)$ can be obtained.
As a result, no exact $V(x(\phi))$ can be obtained from (13).
More complicated forms of metrics with the desired properties
may lead to an exact integrable $\phi^{'}(x)$ and its inverse $x=x(\phi)$,
but (i)-(iii) have the advantage that the outer and inner horizons
can be expressed explicitly in terms of the $A_j$. In addition, (ii) is the
form found in both $2D$ string gravity and $4D$ $GR$ and (iii) is
found in $2D$ string gravity, it is interesting to use them as a
basis of discussions in our present $R=T$ case. As a conclusion,
asymptotically flat double-horizon solutions (i)-(iii)
can be obtained with a presence of a dilaton but no closed
and invertible forms of the dilaton (and $V(\phi)$) can be found.

We now consider the forms (i)-(iii) in the presense of a non-trivial
$\psi(x)$ and $l(\phi)$. Before we do so, we need to know the
equations to calculate the energy and mass for a static solution.
General discussions on the concepts of quasilocal energy and mass in three
and four dimensions can be found in [27]. More recently a general formula for
the
quasilocal energy and mass in two dimensions for an arbitrary $2D$ gravity
theory
has been derived [28]. For the theory described by (2),
using the coordinates (10) we have
$$
\varepsilon(x) = {1\over\sqrt\lambda}(\lambda^{'}+C)-\varepsilon_o, \eqno(21)
$$
for the quasilocal energy evaluated at a spatial position $x$ and
$$
{\cal M}(x) =(\lambda^{'}+C)-{\cal M}_o, \eqno(22)
$$
for the mass. Here $C$ is the integration constant in (15), which is determined
by $T_{\mu\nu}$ in (4) and (8), and ${\cal M}_o$ and $\varepsilon_o$ are the
mass and
energy for a chosen background static spacetime $\lambda_o$.
One can see that the energy $\varepsilon={{\cal M}\over\sqrt\lambda}$
($\varepsilon_o={{\cal M}_o\over\sqrt{\lambda_o}}$) is simply the mass
appropriately blueshifted.
For an asymptotically flat spacetime one can take the limit $x\rightarrow
\infty$ in (22)
to find the total mass $m \equiv {\cal M}(\infty)$ associated with a given
static spacetime. We shall use
(22) to calculate the mass of the static solutions we obtain. Note that if the
spatial
co-ordinate $x$ is replaced by $r\equiv\vert{x}\vert$, then every static
solution
in terms of $r$ remains a solution provided that either (i) it is considered to
be a
``dilaton-spectator vacuum'' solution outside of a distribution of additional
matter stress energy
(such as a perfect fluid with pressure) or (ii) an appropriate delta-function
point-source of
stress energy is inserted at the origin [26]. From now on, we will
express every static solution in terms of $r$. This choice is spatially
symmetric and renders the solutions somewhat more analogous to the four
dimensional
spherically symmetric cases. In the following we consider four classes of
examples, (3.1), (3.2), (3.3) and (3.4) where they
admit asymptotically flat black hole
solutions with double (or multiple) horizons, and one special black hole
solution
which has only one event horizon but admits an infinite chain of universes
connected by timelike wormholes.
Note that in all the examples, $C$ and $C_i$ will denote coupling
constants which appear as integration constants in action (2).

\vskip 0.3 true cm

\noindent{({\sl{3.1}}) $\lambda(r)=\pm{1}+{\beta^{3}\Lambda\over r}+C_2r$,
$\phi=-ln\bigl({r\over\beta}\bigr)$.

\vskip 0.3 true cm

With this form of the metric and dilaton, equation (13) is satisfied if
$$
V(\phi)=-2{\Lambda}e^{3\phi} \eqno(23a)
$$
and (18) yields
$$
{C_0^2\over l(\phi)}={3C_2\over\beta}e^{\phi}+
{3\over 2\beta^2}e^{2\phi}+{\Lambda}e^{3\phi},
$$
$$
b=-{3\over 4}, \quad C_1=-2C_2^2 \eqno(23b)
$$
where we have set $b=-{3\over 4}$ and $C_1=-2C_2^2$ in (18) as an
example. The equation $\lambda(r)=0$ can at most admit two real
positive roots. Note that for the generalized metric
$\pm{1}+{\beta^{n+2}\Lambda\over r^n}+C_2r$, a graphical analysis
shows that $\lambda(r)=0$ still has at most two real positive roots
for all $n>0$. Recall that $C$ and $C_i$, $i=0,1,2...$
denote the coupling constants in the action (2) on which
to constants of integration in the solutions we obtain functionally depend.
The rest of the constants, $\beta$ and $\Lambda$ are both coupling constants.
In general, $\psi^{'}(r)$ in (19) cannot be integrated exactly except for the
choice of $C_1$ in (23b), in which case
$$
C_0\psi=-{1\over 4L}ln\Bigl(r_o\Bigl(-{1\over r}+{L\over r^2}
+C_2\Bigr)\Bigr)+{1\over r}+
{1-8LC_2\over 2L(4LC_2-1)^{1\over 2}}
arctan\Bigl({-1+2C_2r\over -1+4LC_2}\Bigr), \eqno(23c)
$$
where $L=\beta^3\Lambda$ and $-4LC_0\psi_o=ln(r_o)$ is an integration constant.
One can see that as $\lambda(r)\rightarrow 0$, $\psi(r)$ diverges, as we have
mentioned this fact earlier.
The existence of event horizon(s) depends on the relative
magnitude and signs of $C_2$ and $\Lambda$.
Therefore we have to determine the mass parameter
in $\lambda(r)$ first. Naively, one expects $C_2$ to be
related linearly to the mass parameter since in the
vacuum case ($\Lambda=\phi=\psi=0$),
the solution has the form $\lambda=\pm{1}+C_2r$, for
solution has $C_2$ as the mass parameter [26,28].
So we will choose the background mass as the mass of the
reference spacetime with $C_2=0$. It is straightforward but
lengthy to show that (4), (8), (15) and (23b) together imply
$$
C=\pm\sqrt{-2C_1}=\pm{2C_2}. \eqno(23d)
$$
Now (22) yields
$$
m=-C_2, \qquad or \qquad m=3C_2 \eqno(23e)
$$
for the total mass as seen at infinity.
If $\Lambda=0$ in (23b), then $\lambda=\pm{1}+C_2r$
similar to the vacuum case discussed in [2,26,28].
However, $C_2=2m$ in the vacuum case (see (20) with $q^2=0$) rather than
the relation in (23e). Therefore $C_2$ is ``rescaled'' or
flips sign due to the nontrivial presence of $l(\phi)$ in (23b).

Since $m$ is related to $C_1$ ($=-2C_2^2$) through (23e),
the black hole solution admitted by  $l(\phi)$ in (23b)
is a rather special case of the full range of solutions associated with
this coupling function. Semi-classically when the black hole solution radiates,
we expect the mass will change; as a consequence, one can accomodate this by
either modifying the action (2) so that its coupling constants vary with time,
or by performing a full semiclassical analysis of the action (2) with arbitrary
fixed
coupling constants to determine the complete evolution of the evaporating black
hole
whose mass is initially given in terms of these constants via (23e).
Note that for a general $C_1$, the mass is given by
$$
m=\pm\sqrt{-2C_1}+C_2. \eqno(23f)
$$
It is obvious that if $C_2=\pm\sqrt{-2C_1}$, then one obtains an ``massless''
black hole solution provided that the delta-function point-source
at the origin is removed ($r=\vert{x}\vert\rightarrow x$ in the solution).
Note that regardless of whether or not the black hole has mass,
there is a curvature singularity at the $x=0$ due to the bad behaviour of the
energy momentum tensor (8) there (${T^{\nu}}_{\nu}$
or the energy density of (8) diverges at $x=0$).
This massless black hole has an advantage that $l(\phi)$ is no longer
depends on any integration constant. However, it is
not clear how to do thermodynamical analysis on it since it has no
mass. We will use (23e) in the following.

We demand that $m$ in (23e) is always positive.
The quasilocal energy is given by (21).
It approaches the mass as $r\rightarrow\infty$.
Substituting (23f) into the metric yields either
$$
\lambda=\pm{1}+{\beta^3\Lambda\over r}-mr,
\qquad or \qquad \lambda = \pm{1}+{\beta^3\Lambda\over r}+{m\over 3}r.
\eqno(23g)
$$
Due to the arbitrary sign of $\Lambda$,
we expect that (23g) generally admits both event and cosmological
horizons. There are several cases of interest:

\vskip 0.3 true cm

\noindent{(a): $\lambda=1-{L\over r}+{m\over 3}r$}

\vskip 0.3 true cm

We have set $-\beta^3\Lambda\equiv L>0$.
Since $m$ is positive, ${m\over 3}>0$. A simple graphical analysis
shows that there exists only one event horizon in this case.
As $r\rightarrow +\infty$, $\lambda\rightarrow +\infty$. As
$r\rightarrow 0^{+}$, $\lambda\rightarrow -\infty$.
$\lambda^{'}={L\over r}+{m\over 3}$ which is never
zero. Since $\lambda(r)$ is a continous function of $r$, it is
easy to see that only one event horizon exists. The
spatial co-ordinate of the event horizon is given by
$$
r_h={3\over 2m}\Bigl(-1+\sqrt{1+{4mL\over 3}}\Bigr). \eqno(23h)
$$

\vskip 0.3 true cm

\noindent{(b): $\lambda=1-{L\over r}-mr$}

\vskip 0.3 true cm

As $r\rightarrow +\infty$, $\lambda\rightarrow -\infty$. As
$r\rightarrow 0^{+}$, $\lambda\rightarrow -\infty$.
$\lambda^{'}={L\over r^2}-{m}$ thus there exists
one turning point. From the above information, it is easy to
conclude that one may have one event horizon and one
cosmological horizon. Both horizons are given by
$$
r_{\pm}={1\over 2m}(1\pm\sqrt{1-4mL}), \eqno(23i)
$$
where $+$ refers to the cosmological horizon and $-$ refers to
the event horizon.

\vskip 0.3 true cm

\noindent{(c): $\lambda=-1+{L\over r}+{m\over 3}r$}

\vskip 0.3 true cm

In this case we set and $L=\beta^3\Lambda>0$.
Now, as $r\rightarrow +\infty$, $\lambda\rightarrow +\infty$.
As $r\rightarrow 0^{+}$, $\lambda\rightarrow +\infty$. Also,
$\lambda^{'}=-{L\over r^{2}}+{m\over 3}$
which implies there exists one turning point. $\lambda(r)$ is a parabola-like
curve opening upward. Therefore it may have inner and outer horizons
given by
$$
r_{\pm}={3\over 2m}\Bigl(1\pm\sqrt{1-{4mL\over 3}}\Bigr). \eqno(23j)
$$
When $1={4mL\over 3}$, one gets an extremal case. If $1<{4mL\over 3}$, then
the singularity at $r=0$ will be naked.

\vskip 0.3 true cm

\noindent{(d): $\lambda=-1-{L\over r}+{m\over 3}r$}

\vskip 0.3 true cm

This is the final case of the form (3.1) we discuss now. As
$r\rightarrow +\infty$ and $0^{+}$, $\lambda\rightarrow +\infty$
and $-\infty$ respectively. In addition, $\lambda^{'}$ never becomes
zero and so there exists exactly one event horizon which is located at
$$
r_h={3\over 2m}\Bigl(1+\sqrt{1+{4mL\over 3}}\Bigr). \eqno(23k)
$$
We will discuss the thermodynamics properties of the particular case
(c) in the next section.

\vskip 0.3 true cm

\noindent{({\sl{3.2}}) $\lambda=1-{\Lambda_1\over K^2}e^{-Kr}-{\Lambda_2\over
4K^2}e^{-2Kr}$
, $\phi=-{K\over 2}r$, $K>0$}

\vskip 0.3 true cm

This form of solution was first obtained in a $2D$ string theory [24].
For appropriate choices of magnitude and signs of $\Lambda_1$ and $\Lambda_2$,
this asymptotically flat metric admits black hole solutions.
$\lambda(r)$ has no curvature singularity at the origin apart from
a delta-function singularity provided that $\lambda^{'}(0)\neq 0$.
Now, this solution gives $b=-1$,
$$
V(\phi)=\Lambda_1e^{2\phi}+\Lambda_2e^{4\phi}, \eqno(24a)
$$
and
$$
{C_0^2\over l}=C_1+{K^2\over 2}+{3\over 4}\Lambda_2e^{4\phi}-{1\over
2K^2}\Lambda_1\Lambda_2e^{6\phi}
-{3\over 32K^2}\Lambda_2^2e^{8\phi}. \eqno(24b)
$$
$\psi(r)$ in (19) is calculated to be
$$
C_0\psi=-K^2\Bigl({3\over 2}-{2q_1^2\over q_2^2}\Bigr)r
+\Bigl(\Bigl({3\over 4}-{q_1^2\over q_2^2}\Bigr)+{1\over 2}\Bigl(C_1+{K^2\over
2}\Bigr){1\over K}\Bigr)
ln(e^{2Kr}-2q_1e^{Kr}+q_2^2)
$$
$$
+Kq_1\Bigl(q_2^2\Bigl(C_1+{K^2\over 2}\Bigr)-\Bigl({5\over 2}-{2q_1^2\over
q_2^2}\Bigr)(q_2^2-q_1^2)^{-{1\over 2}}
\Bigr)arctan\Bigl({e^{Kr}-q_1\over (q_2^2-q_1^2)^{1\over 2}}\Bigr),
 \eqno(24c)
$$
where we have set $q_1={\Lambda_1\over 2K^2}$ and
$q_2^2=-{\Lambda_2\over 4K^2}$. (4), (8), (15) and (24b) together imply
$$
C=\pm\sqrt{-2C_1}. \eqno(24d)
$$
$C$ has the same dependence on $C_1$ as in the previous
case (3.1). Note that when $C_0=0$ and $\Lambda_2=0$, one
gets back the black hole solution obtained in [26] but with
$C=K$; the mass may be shown to be equal to $K$ when the background
mass ${\cal M}_o$ is chosen to be zero [28]. As the solution (3.2) contains
the black hole solution in [26] as a special case, we set ${\sqrt{-2C_1}}=K$ in
(24b)
and choose ${\cal M}_o=0$ again. Now from (22) we get
$$
m = K={\sqrt{-2C_1}}. \eqno(24e)
$$
Since the mass $m$ is positive, $\lambda(r)$ is asymptotically flat.
$C_0$, $C_1$ ($=-{K^2\over 2}$) and $K$ are
all coupling constants which appear as integration
constants in (24b).
$\Lambda_1$ and $\Lambda_2$ are coupling
constants. Now when $\Lambda_1>0$ and $\Lambda_2<0$,
$\lambda(r)$ has an outer and an inner
horizon. Their spatial co-ordinates are given by
$$
r_{\pm}={1\over m}ln\Bigl(1\pm\sqrt{1+{m^2\Lambda_2\over\Lambda_1^2}}\Bigr)
+{1\over m}ln{\Lambda_1\over 2m^2}. \eqno(24f)
$$
The extremal black hole is obtained when $\Lambda_1^2=-m^2\Lambda_2$.
By rescaling the co-ordinates $r\rightarrow{{\tilde {l}}\over m}r$ and
$t\rightarrow{m\over{\tilde {l}}}t$, one can take the limit $m\rightarrow 0$.
The metric now is $\lambda=-({\Lambda_1\over{\tilde {l}}^2}e^{-{\tilde {l}}r}
+{\Lambda_2\over 4{\tilde{l}}^2}e^{-2{\tilde{l}r}})$ and $\phi={\tilde {l}}r$.
Since $\Lambda_1>0$ and $\Lambda_2<0$, the event horizon disppears but a
cosmological horizon emerges.

In string-inspired theories, the mass parameter appears as a coefficient
${2m\over K}$ in front of the term $e^{-Kr}$ in the metric,
and the square of the electric charge appears
linearly in front of $e^{-2Kr}$ [24,26], $K$ being
related to a coupling constant in the string theories instead of
a coupling constant that determines an integration constant.
In the present case, $\Lambda_2$ can be interpreted as the square of a charge
(recall that in (12), the Maxwell contribution is absorbed into the
second term of $V(\phi)$); therefore only $\Lambda_1$ in $V(\phi)$
is the coupling constant. However, the mass parameter is given by $K=m$ (see
(24e)).
This fact leads to significantly different black hole thermodynamic behaviour.
It is obvious that one may adjust $C_1$ in (22), (24b) and (24d)
in order to make the mass parameter appear
linearly in front of $e^{-Kr}$ in the metric as in the string case in [24].

Since $C_1$ is the coupling in (2) that determines the
integration constant, we have the freedom to ``fix''
it. In particular, if one sets $\sqrt{-2C_1}={\Lambda_1\over\tilde{L}}$
(which implies $\Lambda_1=\tilde{L}m$, see (24e)), where $\tilde{L}>0$ and has
the
same dimension as $m$ and is a coupling constant, then
the mass is now linear in front of $e^{-Kr}$. More precisely,
the metric is now given by
$$
\lambda=1-{m\tilde{L}\over K^2}e^{-Kr}-{\Lambda_2\over 4K^2}e^{-2Kr}.
\eqno(24g)
$$
In this case the mass parameter depends on both  $C_1$ and $\Lambda_1$, the
latter two
quantities  no longer being independent.  $K$ now is a coupling constant.
In the present situation not only $l(\phi)$ explicitly depends on $m$
but also $V(\phi)$.
In the presence of a non-trivial $l(\phi)$, the ``position'' of the
mass in the metric can be adjusted by varying $C_1$ in $l(\phi)$.
It is not surprising since different $C_1$'s corresponding to different
theories
for action (2).
We see that in order for the $R=T$ theory to ``reproduce'' the results
in string theory [24], not only are an additional scalar $\psi$ and a coupling
function $l(\phi)$  required in the matter action, but the constants
of integration in the solution are functions of the coupling constants
in the action.
Finally, if we set $C_1=0$, for example,
then we will get a massless black hole metric provided that
$r\rightarrow x$ ({\sl{i.e.}} the delta-function point-source is removed).
In this situation, $l(\phi)$ and $V(\phi)$ will no longer
depend on the mass.

\vskip 0.3 true cm

\noindent{({\sl{3.3}) $\lambda=1-{\beta^3\Lambda_1\over
2r}-{\beta^4\Lambda_2\over 6r^2}$, $\phi=-ln\bigl({r\over\beta}\bigr)$}

\vskip 0.3 true cm

The choice of these forms for $\lambda(r)$ and
$\phi(r)$ is particularly interesting
since the same form is obtained in a modified version of
$2D$ string theory [29]. In $4D$ general relativity,
the Reissner-Nordstrom solution also has the same metric form.
Now (13) is satisfied if
$$
V(\phi)=\Lambda_1e^{3\phi}+\Lambda_2e^{4\phi}  \eqno(25a)
$$
whereas (18) implies that
$$
{C_0^2\over l(\phi)}={1\over 6}\Lambda_1\Lambda_2\beta^2 e^{5\phi}
+\Bigl(-{1\over 3}\Lambda_2+{5\over 8}\Lambda_1^2\beta^2\Bigr)e^{4\phi}
-3\Lambda_1 e^{3\phi}+{4\over\beta^2}e^{2\phi}+C_1,
\eqno(25b)
$$
where we have set $b=-2$ in (19) for simplicity (note that an extra term
$e^{6\phi}$ will appear in $l(\phi)$ when $b\neq -2$).
$\psi^{'}(r)$ can be integrated exactly and so
$$
C_0\psi=\Bigl({2q_1(q_2^2-q_1^2)\over q_2^4}+C_1q_1\Bigr)
ln\Bigl(1-{2q_1\over r}+{q_2^2\over
r^2}\Bigr)+2C_1q_1ln\bigl({r\over\beta}\bigr)
-{2(q_1^2+q_2^2)\over q_2^2}{1\over r}+{2q_1\over r^2}
$$
$$
+C_1r+{2q_1^2-q_2^2\over (q_2^2-q_1^2)^{1\over 2}}\Bigl(C_1-{2\over
q_2^4}(q_2^2-2q_1^2)\Bigr)
arctan\Bigl({r-q_1\over (q_2^2-q_1^2)^{1\over 2}}\Bigr),
\eqno(25c)
$$
where for simplicity we have set $q_1={\beta^3\Lambda_1\over 4}$
and $q_2^2=-{\beta^4\Lambda_2\over 6}$. Now, one can show that
$$
C=\pm\sqrt{-2C_1}, \eqno(25d)
$$
in (15) by using (4), (8) and (25b). Here $C$ has the same dependence
on $C_1$ as in the previous cases (3.1) and (3.2).
As previously mentioned, different $C_1$'s
correspond to different choices of theory.
Similar to the previous case, we set ${\cal M}_o=0$.
We further set $\sqrt{-2C_1}={\Lambda_1\over 4{\tilde{L}}_1}$,
where ${\tilde{L}}_1>0$ and is a coupling constant with a dimension
of inverse length. This choice for $C_1$ is interesting since it yields
a Reissner-Nordstrom metric form (25f) with the same thermodynamic properties.
Now $C_0$, $C_1$ and $\Lambda_1$ are no longer independent, and the
integration constant $m$ depends on the remaining independent combination of
these. ${\Lambda_2\over 6}$ can be interpreted as the
square of electric charge (see (25f)). Therefore only $\beta$ and ${\tilde{L}}$
are the only remaining independent coupling constants. Using (25d), where the
positive sign is being used, we obtain
$$
m={\Lambda_1\over 4\tilde{L}_1}. \eqno(25e)
$$
The metric is now given by
$$
\lambda=1-{2\beta^3{\tilde{L}}_1 m\over r}-{\beta^4\Lambda_2\over 6r^2}.
\eqno(25f)
$$
Now the metric is exactly the Reissner-Nordstrom metric (with the same
thermodynamics properties) obtained in $2D$ string theory and $4D$ $GR$,

Note that the presence of $\beta^3$ and $\beta^4$ in the
$r^{-1}$ and $r^{-2}$ terms are due to the fact that
mass and charge both have dimensions of inverse length in
two dimensions, instead of length in four dimensions.
As a result, extra length scales $\beta^3$ and $\beta^4$
are required. If $\Lambda_2<0$ in (25f), then we have outer and
inner horizons. We comment that $V(\phi)$ and $l(\phi)$ both
explicitly depend on the mass parameter $m$ similar to
the situation in the last section. However, if one sets
$C_1=0$ instead, one will get a massless black hole.

\vskip 0.3 true cm

\noindent {({\sl{3.4}}) Multiple Horizons}

\vskip 0.3 true cm

In this section, we illustrate two examples which admit multiple-horizon
spacetime structures. Multiple-horizon structures are
possible in two-dimensional gravity due to simpler field
equations. Our first example is
$$
\lambda=1-{\Lambda_1\over K^2}e^{-Kr}-{\Lambda_2\over 4K^2}e^{-2Kr}
-{\Lambda_3\over 9K^2}e^{-3Kr}, \qquad \phi=-{K\over 2}r. \eqno(26a)
$$
This form of $\lambda(r)$ and $\phi(r)$ was obtained in [24,29].
Now (13), (18) and (26a) yield
$$
V=\Lambda_1 e^{2\phi}+\Lambda_2 e^{4\phi}+\Lambda_3 e^{6\phi},
$$
$$
{C_0^2\over l}=C_1+{K^2\over 2}+{3\over 4}\Lambda_2 e^{4\phi}+
\Bigl(-{1\over 2K^2}\Lambda_1\Lambda_2+{8\over 9}\Lambda_3\Bigr)e^{6\phi}
$$
$$
-{1\over K^2}\Bigl({3\over 32}\Lambda_2^2+{2\over
3}\Lambda_1\Lambda_3\Bigr)e^{8\phi}
-{1\over 6K^2}\Lambda_2\Lambda_3 e^{10\phi}-{4\over 81K^2}\Lambda_3^2
e^{12\phi}
\eqno(26b)
$$
with $b=-1$. Generally speaking, one can have $1+\sum_{n=1}e^{-nKr}$
in (26a) by adjusting $V(\phi)$ and $l(\phi)$ in (13) and (18) respectively.
For simplicity, we just illustrate the first three terms.
Similar to case (3.2), where only the first two terms were considered,
$C$ is still given by (24d).
We again set $\sqrt{-2C_1}={\Lambda_1\over\tilde{L}}$ in (26b)
and obtain exactly the three-horizon black hole obtained in string
theory [24,29]. If one instead sets $C_1=0$,
one gets a massless black hole. For the former choice
$C_0$, $C_1$ and $\Lambda_1$ are
all coupling constants
in the action (2), on which the integration constants for the
solution (26a) functionally depend. $\Lambda_2$ can be interpreted as square of
an electric
charge, and $K$, $\tilde{L}$ and $\Lambda_3$ are coupling constants.

Regardless of whether we choose a massless
black hole or not, it can be shown that
$$
e^{-Kr_i}=-{3\over 4}{\Lambda_2\over\Lambda_3}\Bigl(1-2\sqrt{1-{16\over
3}{\Lambda_1\Lambda_3\over\Lambda_2^2}}cos\theta_i\Bigr)
$$
$$
cos(3\theta_i)=-\Bigl(1-{16\over
3}{\Lambda_1\Lambda_3\over\Lambda_2^2}\Bigr)^{-{3\over 2}}
\Bigl(1-8{\Lambda_1\Lambda_3\over\Lambda_2^2}-{32\over
3}{\Lambda_3^2\over\Lambda_2^3}K^2\Bigr).
\eqno(26c)
$$
Here $r_1$, $r_2$ and $r_3$ are the three possible roots of the
cubic equation $\lambda(r)=0$ in (26a). When $\Lambda_1=0$,
(26c) indicates that there exists three real roots
provided that $0<{K^2\Lambda_3^2\over\Lambda_2^3}\leq {3\over 64}$.
When all $\Lambda_1$, $\Lambda_2$, $\Lambda_3$ are non-vanishing,
a graphical analysis gives us the following conditions on
$\Lambda_1$, $\Lambda_2$ and $\Lambda_3$ for the cubic equation $\lambda(r)=0$
to
admit three positive real roots. As $r\rightarrow\infty$, $\lambda\rightarrow
1$.
If $K^2<\Lambda_1+{\Lambda_2\over 4}+{\Lambda_3\over 9}$,
then $\lambda<0$ as $r\rightarrow 0^{+}$.
In addition, from the equation $\lambda^{'}=0$ and the continuity of $\lambda$,
one can check that there are two turning points for $\lambda(r)$
(one is the maxima, the other one is the minima) provided that
$1>{16\over 3}{\Lambda_1\Lambda_3\over\Lambda_2^2}>0$, $\Lambda_1>0$
and $\Lambda_2<0$. The above conditions on $\Lambda_i$ yield three horizons.
The Penrose diagram of such black hole
with three horizons is called the $2D$ lattice shown in [24].

Another example in this section
is the ``extension'' of the Reissner-Nordstrom solution.
It is given by
$$
\lambda=1-{\beta^3\Lambda_1\over 2}{1\over r}
-{\beta^4\Lambda_2\over 6}{1\over r^2}
-{\beta^5\Lambda_3\over 12}{1\over r^3}, \qquad
\phi=-ln\bigl({r\over\beta}\bigr). \eqno(26d)
$$
This form of $\lambda(r)$ and $\phi(r)$ can be found in [24,29].
As usual, $V(\phi)$ and $l(\phi)$ can be determined from the given $\lambda(r)$
and the invertible $\phi(r)$. For (26d), (13) and (18) show that
for $b=-2$ as in section (3.3), they are given by
$$
V=\Lambda_1 e^{3\phi}+\Lambda_2 e^{4\phi}+\Lambda_3 e^{5\phi},
$$
$$
{C_0^2\over l}=C_1+{4\over\beta^2}e^{2\phi}-3\Lambda_1 e^{3\phi}
+\Bigl({5\over 8}\beta^2\Lambda_1^2-{1\over 3}\Lambda_2\Bigr)e^{4\phi}+
\Bigl({1\over 6}\beta^2\Lambda_1\Lambda_2+{1\over 3}\Lambda_3\Bigr)e^{5\phi}
$$
$$
-{1\over 8}\beta^2\Lambda_1\Lambda_3 e^{6\phi}-{1\over
18}\beta^2\Lambda_2\Lambda_3 e^{7\phi}
-{7\over 288}\beta^2\Lambda_3^2 e^{8\phi}. \eqno(26e)
$$
More terms ($r^{-4}$, $r^{-5}$...) can be added to (26d) by adjusting
$V(\phi)$ and $l(\phi)$. $C$ is still given by
$C=\pm\sqrt{-2C_1}$, and one can have the mass parameter related to $\Lambda_1$
as in (25e) in case (3.3). Now we obtain the
three-horizon black hole obtained in [29]. The three possible
roots for $\lambda=0$ in (26d) are given by
$$
{1\over r_i}=-{2\over 3}{1\over\beta}{\Lambda_2\over\Lambda_3}
\Bigl(1-2\sqrt{1-{9\over
2}{\Lambda_1\Lambda_3\over\Lambda_2^2}}cos\theta_i\Bigr)
$$
$$
cos3\theta_i=-\Bigl(1-{9\over
2}{\Lambda_1\Lambda_3\over\Lambda_2^2}\Bigr)^{-{3\over 2}}
\Bigl(1-{27\over 4}{\Lambda_1\Lambda_3\over\Lambda_2^2}-{81\over 4\beta^2}
{\Lambda_3^2\over\Lambda_2^3}\Bigr).
\eqno(26f)
$$
If $\Lambda_1=0$, then (26f) indicates that there exist three positive roots
provided that $0<{\Lambda_3^2\over\Lambda_2^3}{1\over\beta^2}\leq {2\over 81}$
is satisfied.
If all $\Lambda_1$, $\Lambda_2$ and $\Lambda_3$ are non-vanishing, a
simple grapical analysis shows that
there exist three real positive roots provided that
$1>{9\over 2}{\Lambda_1\Lambda_3\over\Lambda_2^2}>0$, $\Lambda_2<0$,
and $\Lambda_1>0$.

Finally we illustrate an example which has only one event horizon but admits
interesting causal structure. The metric is given by
$$
ds^2=-\Bigl(1-{s\over 1+cosh(Qx)}\Bigr)dt^2+{dx^2\over \Bigl(1-{s\over
1+cosh(Qx)}\Bigr)},
\qquad \infty>s >2. \eqno(27)
$$
This metric was previously obtained in [30,31] for
a finite $s$ (a real number) as a solution
to $2D$ string gravity with a non-zero tachyon field.
Re-writing it in non-Schwarzchildian co-ordinates and taking the
limit $s\rightarrow\infty$ it reduces to the string metric
in (3.2) in the same co-ordinates. It is believed that
it solves the $\beta$-function equations exactly -- this  has been confirmed
by explicit computation to four-loop order (see discussions in [30]).
$Q$ is a coupling constant and has an inverse length.
(27) is symmetric and asymptotically flat as $x\rightarrow\pm\infty$.
Note that unlike previous sections, we have used $x$ instead of
$r=\vert{x}\vert$.
The horizon is located at $Qx_H=cosh^{-1}(s-1)$. Note that the
dilaton for the solution (27) in [30] or [31] is not invertible. Since the
classical geometry of the spacetime is solely determined by the metric (27)
and is not affected by the form of dilaton, we will choose an
invertible dilaton, $\phi(x)=ln(cosh(Qx)+1)$, which simplifies the
calculations,
in order to obtain $l(\phi)$ and $V(\phi)$.
With that form of dilaton, the potential and coupling functions are
$$
V=sQ^2(e^{-\phi}-3e^{-2\phi}),
$$
$$
{C_0^2\over l}=Q^2\Bigl(-(3s+4)e^{-\phi}+s({3\over 2}s+5)e^{-2\phi}
-2s^2e^{-3\phi}+{C_1\over Q^2}+2\Bigr). \eqno(28)
$$
$C$ in (15) is again given by
$$
C=\sqrt{-2C_1}. \eqno(29)
$$
If we set ${2{\tilde{L}}^2\over s}=-2C_1$, where $\tilde{L}$ is an
integration constant with an inverse length, and choose ${\cal M}_o=0$, then
(22)
yields the $ADM$ mass
$$
m=\tilde{L}\sqrt{2\over s}. \eqno(30)
$$
which is exactly the mass formula obtained in [31].
Now $C_0$ and $C_1$ (or $\tilde{L}$) are  coupling constants on which
the integration constants in (28) functionally depend.
Both $R=T$ and $2D$ string gravity (with a tachyon field)
yield the same classical black hole geometry with the same thermodynamic
properties.

If we perform a co-ordinate transformation,
$Qx\rightarrow cosh^{-1}\Bigl({(s-2){{\tilde{x}}\over\beta}+s\over 2}\Bigr)$,
where $\beta$ is a length scale, then (27) becomes
$$
ds^2=2(s-2)\Bigl(-f({\tilde{x}})dt^2+{\beta^2 d{\tilde{x}}^2\over
4({\tilde{x}}^2-\beta^2)}\Bigr),
$$
$$
f(\tilde{x})=\Bigl({\tilde{x}+\beta\over \tilde{x}-\beta}-{2\over
s}\Bigr)^{-1}, \eqno(31)
$$
which was obtained in [32] with $\beta$ scaled to be $1$.
Using the chosen dilaton above we can
obtain $l(\phi)$ and $V(\phi)$ in the $\tilde{x}$ co-ordinate. In fact,
(27) only covers part of the spacetime of (31) [30]. It was shown in
[30] that the maximally extended exact black hole geometry corresponds to
an infinite sequence of asymptotically flat regions linked
by wormholes (it is reminiscent of the causal structure of the
Reissner-Nordstrom
black hole, except that the singularities are removed).
Such an unusual causal structure has no analog in $D>2$ $GR$
but it is possible in the present $R=T$ theory ($D\rightarrow 2$ limit of
$4D$ $GR$).

Before ending this section, we want to comment that the above approach in
constructing exact static solutions is very straightforward
in two spacetime dimensions due to the fact that (10) has only one
degree of freedom and the field equations are much simplier than their
higher dimesnional counterparts. In higher dimensions, such an approach is
possible [19] in the $FRW$ single scalar field cosmology
since the metric only has one dynamical degree
of freedom, namely the scale factor $a(t)$. However,
higher dimensional static spherically symmetric solutions
in any gravitational theory are typically not obtained in the approach adopted
here.
We also note that $l(\phi)$ in all
the above solutions is generally a function of $r$. It may have different
signs for different ranges of $r$ outside (inside) the event (cosmological)
horizon. When $l(\phi)$ is negative (positive), $\psi$ behaves as a
massless scalar (ghost). Although the overall quasilocal energy and mass (or
$ADM$)
are positive and finite for all $r$,
the ambiguity of the transitions $scalar\leftrightarrow ghost$
and the stability of all the above static solutions are open questions.

\vskip 0.6 true cm

\noindent{\bf{4 Equations of Motions of Test Particles}}

\vskip 0.3 true cm

The third term in (19), when performing the integration over
$r$, becomes ${1\over 2}(\lambda^{'}ln\lambda-\int{ln\lambda}d\lambda^{'})$.
At an event or cosmological horizon $\lambda=0$ and so $\psi$
generally diverges at the horizon (see (23c), (24c), (25c) for examples).
One might argue that since the action (2) and its
equations of motion do not explicitly depend on $\psi$, its divergence
may not be physically pathological. However the term $l(\phi)(\nabla\psi)^2$ in
(2)
(which is equal to $C_0\psi^{'}$ from (19) in metric (10) with spatial $r$),
also diverges at the horizons.
In this section, we will concentrate our discussions only on event horizons
and show that a particle with ``spectator charge'' ({\it i.e.} one
linearly coupled to $\psi^{'}$ in the coordinates we use)
will encounter an infinite potential barrier before hitting the event
horizon. In four dimensions the static
``Bekenstein black hole'' considered in [33] is a solution
of coupled Einstein-Maxwell conformal scalar field equation and
encounters a similar problem. The conformal scalar,
which appears explicitly in the action, diverges at the event horizon.
However, it was shown in [33] that the infinity in the scalar field
need not to be physically pathological due to completeness of the
trajectory and the absence of both infinite tidal forces
and an infinite potential barrier at the event horizon for
a test particle linearly coupled to the conformal scalar.
In the $2D$ black hole metrics considered at the end of the last section,
the string dilaton obtained in [30] or [31] also diverges at the event horizon.

In the following, we will
briefly investigate the motion of a test particle linearly coupled to $\psi$
and $\psi^{'}$. For simplicity, in the following
we will restrict ourselves to the
double-horizon metrics in cases (3.2) and (3.3)
considered in the last section.
These two cases have the properties that as $r\rightarrow r_h$,
$\lambda(r)\rightarrow 0$ and $\lambda^{'}(r)$ and $\lambda^{''}(r)$
are both finite. When $r\rightarrow\infty$, $\lambda(r)\rightarrow 1$
and all its derivatives vanish. For the case (3.1), as
$r\rightarrow\infty$, $\lambda(r)$ diverges. However,
a local Rindler transformation may be applied to rewrite
the asymptotic metric in Minkowski form, as we have mentioned
previously. For simplicity, we will use the $r-t$ co-ordinate
in (10) in the following and consider cases (3.2) and (3.3) only.
Every conclusion drawn from cases (3.2) and (3.3) should be
valid for the case (3.1).
We will closely follow the computations done in [33] since
almost all its computations in the $r-t$ co-ordinates for the radial motion
of a test particle are also valid in our two-dimensional cases.

The simplest parameter-invariant
for a particle of a unit rest mass coupled linearly to a function
$F(\psi, l(\phi)(\nabla\psi)^2)$ is
$$
S=-\int (1+{\eta}F)
\Bigl(-g_{\alpha\beta}{dx^{\alpha}\over d\gamma}{dx^{\beta}\over
d\gamma}\Bigr)^{1\over 2}d\gamma, \eqno(32)
$$
where $\eta$ is a coupling constant, $\gamma$ is a parameter and
$x^{\alpha}(\gamma)$
is the trajectory of the particle. The term proportional to $1$ is the action
for a
free particle; that proportional to $\eta F(\psi, l(\phi)(\nabla\psi)^2)$
is the interaction term for a particle with spectator charge.
We will choose $F(\psi, l(\phi)(\nabla\psi)^2)={l(\phi)(\nabla\psi)^2\over
C_0}$
or $\psi$ in our discussion.

Since the action is invariant under a change of parameter $\gamma$,
we are free to impose a condition on $-g_{\mu\nu}U^{\mu}U^{\nu}$,
$U^{\mu}={dx^{\mu}\over d\gamma}$ to fix the choice of $\gamma$. A useful
choice which simplifies the equation of motion is
$$
(1+\eta F)^2=-g_{\mu\nu}U^{\mu}U^{\nu}=\bigl({d\tau\over d\gamma}\bigr)^2,
\eqno(33)
$$
where $\tau$ is the proper time of the test particle.
Now $\gamma$ is an affine parameter. The equation of motion which follows
from variation of $S$ with respect to $x^{\mu}$ now becomes
$$
{D^2x^{\nu}\over d\gamma^2}=-\eta (1+\eta F)\nabla^{\nu}F,
\eqno(34)
$$
where ${D^2x^{\nu}\over d\gamma^2}=U^{\mu}\nabla_{\mu}U^{\nu}$.
The energy $E$ is a constant of the motion and is
given by
$$
E=\lambda\bigl({dt\over d\gamma}\bigr) \eqno(35)
$$
from which (33) becomes
$$
{dr\over d\gamma}=\pm(E^2-\lambda(1+\eta F)^2)^{1\over 2}.
\eqno(36)
$$
Here $+$ and $-$ refer to outgoing and ingoing trajectories respectively.
Using (33) and (36), we get
$$
(1+\eta F){dr\over d\tau}=\pm(E^2-\lambda(1+\eta F)^2)^{1\over 2}. \eqno(37)
$$
Now we set $F=\psi^{'}$ since this is the term in the action which diverges
at the horizon. (19) indicates that as $r\rightarrow\infty$,
$\psi^{'}\rightarrow -{2b\phi^{'2}\over C_0}+{C_1\over C_0}$.
When the test particle starts its journey
from $r\rightarrow\infty$, (36) becomes
$$
{dr\over d\gamma}=-\Bigl(E^2-\Bigl(1+{\eta\over
C_0}\bigl(-2b\phi^{'2}+C_1\bigr)\Bigr)^2\Bigr)^{1\over 2}.
\eqno(38)
$$
In all the black hole solutions $\phi^{'}$ is finite as
$r\rightarrow\infty$. We can always choose some proper value
of ${\eta\over C_0}$ such that the right hand side of (38) is real and
negative for a given $E$. As $r\rightarrow r^{+}_h$, the event horizon,
$\psi^{'}\rightarrow {1\over 2C_0\lambda}(\lambda^{'2}+2C_1)$.
Thus (36) now becomes
$$
{dr\over d\gamma}=-\Bigl(E^2-{\eta^2\over
4C_0^2\lambda}(\lambda^{'2}+2C_1)^2\Bigr)^{1\over 2}.
\eqno(39)
$$
It is obvious that the right hand side of (39) is now complex and diverging.
It is clear that as $r\rightarrow\infty$, the function inside the bracket
of square root of the right hand side of (36)
can be chosen to be positive, while as $r\rightarrow r^{+}_h$,
it must be negative and diverging. Since the function is continuous for
$\infty > r > r_h$, there must exist a point $r_o$ between infinity and the
event
horizon which is a turning point; that is,
$$
{dr\over d\gamma}=0, \qquad r=r_o. \eqno(40)
$$
(40) shows that with respect to the affine parameter $\gamma$,
there is something like an infinite potential
barrier at $r=r_o$. The particle must stop at $r_o$ and
rebound due to the fact that it is not travelling
along a geodesic ({\it i.e.} it is coupled to $\psi^{'}(r)$). The
non-gravitational
force generated by $\psi^{'}(r)$ in the right hand side of (34) (which vanishes
if
$\eta=0$ in which case the trajactory is a geodesic) produces a replusive
effect. The effect is strong enough to prevent the particle from
reaching the horizon. One can set $\psi(r)$ to be a constant to get rid
of this replusive effect but we cannot get a closed form
for the dilaton and its potential.
More precisely, we can still have an asymptotically flat
double-horizon black hole metric but the form of $\phi(r)$ and $V(\phi)$
are not closed; that is, the solution is not exact (see discussions on section
3).

So far our approach has been to couple the particle to quantities which
are diverging at the event horizon and see what happens.
We set $F=\psi^{'}(r)$ (recall that $C_0\psi^{'}=l(\phi)(\nabla\psi)^2$)
in the above. Apart from being the simplest form,
$\psi^{'}(r)$
explicitly appears in the field equations and it diverges at the
horizon. Although the resultant equations of motion of the test particle
do not follow from the field equations, it is natural to consider
$F=\psi^{'}(r)$
from the above point of view.
In addition $\psi(r)$,
although it does not appear
explicitly or implicitly in the field equations, diverges at the
horizon. Hence we consider dependence of $F$ on $\psi$ as well.

If one sets $F(\psi, \psi^{'})=\psi$, the situation changes.
As $r\rightarrow r_h^{+}$, $C\psi\rightarrow\lambda^{'}ln\lambda$. Thus
(36) now becomes
$$
{dr\over d\gamma}\rightarrow-\Bigl(E^2-{\eta^2\over C_0^2}\lambda\lambda^{'2}
(ln\lambda)^2\Bigr)^{1\over 2}. \eqno(41)
$$
Since $\lambda^{'}(r)$ is finite at $r=r_h$,
and $\lambda (ln\lambda)^2\rightarrow 0$
as $r\rightarrow r_h^{+}$, we have
${dr\over d\gamma}\rightarrow -E$ at the horizon.
As $r\rightarrow\infty$, $\psi(r)$ in (24c) and (25c)
diverge (except when $2C_1=-K^2$ in (24c) and $C_1=0$ in (25c)).
However, we can always let the particle starts from a finite spatial
point $r_1$ such that by choosing some proper value of
${\eta\over C_0}$, ${dr\over d\gamma}$ at $r_1$ is negative and real.
(24c) and (25c) indicate that none of the $\psi(r)$
are monotonic decreasing/increasing
functions of $r$. We no longer conclude that ${dr\over d\gamma}$
will become zero at a point $r_o$ between $r_1$ and the event horizon.
It is also not possible to exactly solve the equation ${dr\over d\gamma}=0$
in (36) to get the $r_o$ due to the presence of the logarithm and
arctangent
functions. Despite the lack of information about the existence of
an infinite potential barrier, we shall now show that even if there is no such
barrier, particles linearly coupled to $\psi$ will encounter an
infinite tidal force at the horizon in a finite proper time.

A criterion for a physical singularity is that the relative tidal
acceleration between pairs of nearby trajectories of some
particles should become unbounded as the trajectories
approach the point in question. Consider a family
of trajectories $x^{\alpha}(\gamma, \nu)$ parametrized by
$\gamma$, an affine parameter along the trajectories, and
$\nu$, a parameter labelling the trajectories. The equation of motion
is given by (34). Defining
$y^{\alpha}={\partial{x^{\alpha}}\over\partial{\nu}}$,
which is the separation vector, the relative acceleration is given by
$$
{D^2y^{\mu}\over
d\gamma^2}={R^{\mu}}_{{}\alpha\beta\sigma}U^{\alpha}U^{\beta}y^{\sigma}
+y^{\beta}\nabla_{\beta} N^{\mu}, \eqno(42)
$$
where $N^{\beta}=-\eta(1+\eta F)\nabla^{\beta} F$ from the right hand side
of (34). Equation (42) is the generalization of the geodesic deviation equation
for particles
with an additional force term [33]. Using (33) and (34), (42) becomes
$$
{D^2y^{\mu}\over d\tau^2}={R^{\mu}}_{{}\alpha\beta\sigma}{dx^{\alpha}\over
d\tau}{dx^{\beta}\over d\tau}y^{\sigma}
$$
$$
-\eta(1+\eta F)^{-1}\bigl(y^{\beta}\nabla_{\beta}\nabla^{\mu}F
+\eta(1+\eta F)^{-1}y^{\beta}\nabla_{\beta}F\nabla^{\mu}F
+(\nabla_{\beta}F){dx^{\beta}\over d\tau}{dy^{\mu}\over d\tau}\bigr). \eqno(43)
$$
We need to express the components of ${D^2y^{\mu}\over d\tau^2}$
in an instantaneously comoving inertial frame to calculate the local
relative acceleration. The dyad  ${e^{\mu}}_a$ (where
$a$ is the dyad index) is constructed by taking ${e^{\mu}}_0$ as the
two-velocity of an infalling timelike geodesic in the metric (10), and
taking the other component orthogonal to this basis vector
and parallel transported along the geodesic.
Also, it is required that the dyad be instantaneously comoving with
the moving test particle at $r=r_p$, where $r_1\geq r_p\leq r_h$.
The dyad satisfying the above conditions is given by [33]
$$
{e^{\mu}}_0=\pm{E\over\lambda}(1+\eta F)^{-1}{\delta^{\mu}}_t
-(E^2(1+\eta F)^{-2}-\lambda)^{1\over 2}{\delta^{\mu}}_r,
$$
$$
{e^{\mu}}_1={1\over\lambda}(E^2(1+\eta F)^{-2}-\lambda)^{1\over
2}{\delta^{\mu}}_t
\pm E(1+\eta F)^{-1}{\delta^{\mu}}_r,
\eqno(44)
$$
where ${e^{\mu}}_a{e^{\nu}}_b g_{\mu\nu}=n_{ab}$
and ${e^{\mu}}_0\nabla_{\mu}{e^{\nu}}_b=0$.
This set of basis vectors has the same expresssion as the $r-t$ part of the
tetrads considered in [33] where only radial motion was considered.
The components of the relative acceleration in the
inertial frame are now given by the right hand side of (43) provided that
the indices are dyad indices ($a,b=0,1$), and the contractions are performed
with
the Minkowski metric $n_{ab}$. The components ${R^{a}}_{bcd}$ are bounded since
the
geometries are all well-behaved at the event horizon for the metrics discussed
in the last section. ${dx^{\alpha}\over d\tau}={\delta^{\alpha}}_0$ by
definition
so it is finite as well. Now we need to calculate $(1+\eta F)^{-1}\nabla_a F$
($=(1+\eta F)^{-1}{e^{r}}_aF^{'}$) and
$(1+\eta F)^{-1}\nabla_{a}\nabla_{b}F$
($=(1+\eta F)^{-1}{e^{\mu}}_a{e^{\nu}}_b\nabla_{\mu}\nabla_{\nu}F)$
in (43). When $F=\psi(r)$, using (24c) and (25c) it is lengthy but
straightforward to check that these quantities all diverge as $r\rightarrow
r_h$.
Using (37) we get near the horizon that
$\delta\tau\rightarrow 0$ for two points $r_h+\delta r$ and $r_h$
(except for extremal cases, where an infinite throat exists).
This indicates that an ingoing test particle coupled to $\psi$ will
hit the horizon within a finite proper time provided there is no
infinite potential barrier outside the horizon.

To summarize, a test particle linearly coupled to either $\psi^{'}(r)$
or $\psi(r)$ may encounter an infinite potential barrier
outside the horizon or a infinite tidal force at the horizon
for a asymptotically flat spacetime which admits a double- or
even multiple-horizon structure. However,
one may argue that apart from being the simplest form, the action (32) is an
artifact since it does not follow from any field equation and therefore
the non-geodesic motion and singularities are ``artificial'' (unlike the
analogous action in [33], where for $F=\psi$ (the conformal scalar)
the action is singled out by its conformal invariance
properties due to the conformal invariance of the Einstein-Maxwell-conformal
scalar equations). Since any form of
equation of motion does not follow from the field equations,
one has the freedom to choose a special form
$F(\psi, \psi^{'})$ (still diverging at the horizons)
in which a test particle with spectator charge encounters
no singularities throughout its motion. We have shown that
at least some forms of $F=\psi(r)$, $\psi^{'}(r)$
yield the two kind of infinities mentioned above for a test particle,
and shall not pursue this issue further.

\vskip 0.6 true cm

\noindent{\bf{5 Hawking Temperature and Entropy}}

\vskip 0.3 true cm

An important thermodynamic quantity in a static black hole solution
is the Hawking temperature $T_H$. The existence of a meaningful temperature
presupposes the existence of both an event horizon and a well-defined value
of the surface gravity at the horizon and is given by [25]
$$
4\pi T_H=\lambda^{'}(r_h). \eqno(45)
$$
The entropy $S$ associated with a two-dimensional black hole is not as
straightforwardly obtained as for higher-dimensional cases, as the
event horizon has no area. It may be deduced from the Noether charge technique
[34] or alternatively from the
black hole analog of the thermodynamic equation $dm=TdS$ when all other
integration constants are fixed [25]. So the entropy is given by integrating
the following equation
$$
{\partial S\over\partial m}={1\over T_H}. \eqno(46)
$$
Note that $T_H$ and $S$ are quantities measured at spatial infinity
and (46) does not hold for an extremal black hole in which
$T_H=0$ (the entropy for an extremal black hole and a finite-space formulation
of black hole thermodynamics require seperated investigations).
We are going to calculate $T_H$ and $S$ in several examples.
For the case ({\sl{3.1}}c), with the choice of mass in (23f),
(45) yields the temperature,
$$
4\pi T_H={m\over 3}{1-{4\over 3}Lm+\sqrt{1-{4\over 3}Lm}\over
1-{2\over 3}Lm+\sqrt{1-{4\over 3}Lm}}. \eqno(47)
$$
Recall that $L=\beta^3\Lambda>0$ in (3.1c).
For the extremal case where $4Lm=3$, $T_H=0$. When
$L=0$, (47) reduces to the equation obtained in [25] for a vacuum black hole
(apart from the factor in front of $m$).
The entropy in (46) can be calculated by integrating (47):
$$
S=12\pi\Bigl(ln\Bigl({2L\over L_o(\sqrt{1-{4\over 3}Lm}+1)}\Bigr)
+ln\bigl({m\over M_o}\bigr)\Bigr), \eqno(48)
$$
where $S_o=-ln(L_oM_o)$ which is an integration constant.
We note that in (48) above, the second term is the entropy contribution
from a vacuum black hole in [25]. The first term in (48) can be
interpreted as the entropy contribution from the potential (23a).

For the case (3.2), (45) implies that
$$
4\pi T_H={2m\sqrt{1+{\Lambda_2m^2\over\Lambda_1^2}}\over
1+\sqrt{1+{\Lambda_2m^2\over\Lambda_1^2}}}, \eqno(49)
$$
where the choice of mass (24e) is used.
Again $T_H=0$ for the extremal case where $\Lambda_1^2=-\Lambda_2m^2$.
When $\Lambda_2=0$ ($\Lambda_2=0$ and (24e) imply $l(\phi)=0$ in (24b)),
(49) reduces to the relation obtained
in [26] for a Liouville field coupled to gravity.
Therefore the thermodynamics of the Louville black hole in [26] is the same as
the vacuum
black hole obtained in [25]. When $l(\phi)$ and $\psi(r)$ are non vanishing,
$T_H$ will be altered according to (49). Now the entropy in (46) is
$$
S=2\pi ln\Bigl(1-\sqrt{1+{\Lambda_2m^2\over\Lambda_1^2}}\Bigr)+S_o. \eqno(50)
$$
When $\Lambda_2\rightarrow 0$, $S\rightarrow 2\pi
ln\bigl(-{\Lambda_2m^2\over 2\Lambda_1^2}\bigr)+S_o$ which can further be
written
as $S\rightarrow 4\pi ln\bigl({m\over M_o}\bigr)$ where the contribution
from vanishing $\Lambda_2$ has been absorbed into $S_o$.
Therefore the entropy reduces
to the case considered in [26]. For the Reissner-Nordstrom
case in ({\sl{3.3}}), the thermodynamic
quantities $T_H$ and $S$ were widely studied. We will not repeat the
calculations here.

Finally, we comment that by adjusting $C_1$
in (18) ({\sl{i.e.}}, adjusting the theory for action (2)),
one can construct alternate thermodynamic behaviours
for the black hole solutions in section 3. As the final example,
we consider metric (24g) with $\Lambda_2=0$
in case (3.2). Using (45) we get the temperature
$$
T_H=K. \eqno(52)
$$
$T_H$ is independent of the mass. This result was previously obtained
in $2D$ string black hole solutions. One sees that with the presence
of $\psi(r)$ and $l(\phi)$, we can have different black hole
solutions with different thermodynamical properties.
Other examples of interest can be constructed.
We will not discuss them further.

\par\vfil\eject

\noindent{\bf{6 $FRW$ Cosmology}}

\vskip 0.3 true cm

In this section, we consider non-static solutions to action (2), namely
the $FRW$-type cosmological solutions. $2D$ $FRW$ cosmological
models were previously considered in [10] for the $R=T$ theory
with the matter action replaced by a general perfect fluid with
the equation of state $p=(\gamma-1)\mu$, where $p$ is the pressure
and $\mu$ is the energy density. In addition, cosmological singularities
were also discussed. It was shown that a singularity in the energy density
can occur in both $FRW$ and $tilted$ models if certain
energy conditions for $T_{\mu\nu}$ are satisfied. Exact scalar
field cosmologies which arise from the action (2) when there are no $\psi$
and $F_{\mu\nu}$ fields in the matter action [35] have also been studied.
A series of exact solutions were derived, including a
non-singular one. This non-singular solution is possible since
in the presence of a scalar field, certain energy conditions
are violated. $R=T$ $FRW$ cosmology can also be a useful model
for studying phase transtions and topological defects in two dimensions,
and such investigations may turn out be important for the
understanding of some of the distinct features of $2D$ gravity [36].

To this end, we study $FRW$ cosmologies which follow from the
action (2) when there are non-trivial
$\psi$ and $l(\phi)$. Similar to the static case, it will be shown that
there are $l(\phi)$ and $V(\phi)$ which can lead to almost any
desired behaviour for the metric. That means one can find the
``cosmological fluid'', which is described by $\phi$ and $\psi$
in the matter action in (2), required to give a desired cosmological behaviour.

For a $2D$ $FRW$ cosmology, the non-static metric is given by [7,10],
$$
ds^2=-dt^2+a^2(t)dx^2, \eqno(53)
$$
where $a(t)$ is the cosmic scale factor. In virtue of (53),
(3), (4) and (8), we get
$$
2{\ddot a\over a}=V. \eqno(54)
$$
The dot denotes an ordinary derivative with respect to
time $t$. Following similar algebraic procedures to the static case, (5)
becomes
$$
{C_0^2\over la^2}=-{C_2\over a^2}-2b{\dot\phi}^2-2\bigl({\dot a\over
a}\bigr)^{.}, \eqno(55)
$$
where we have used the fact that (6) is given by
$$
\dot\psi={C_0\over la}. \eqno(56)
$$
$C_0$ is an integration constant. It is just a length scale with
a dimension of inverse length with no physical significance.
(55) is in fact the $2D$ version of Friedmann equation. In $4D$ $FRW$
cosmology, the analogous constant $C_2$ can be scaled to $0$,
$\pm 1$ corresponding to open flat space,
closed sphere and closed hyperboloid respectively. However from (53) we see
that
the openess/closedness of spacetime has no effect on the evolution of the scale
factor. In fact, $C_2$ has no relationship with the one-dimensional
spatial geometry [10]. On the other hand, $C_2$ in $4D$ $FRW$ cosmology
is related to the intrinsic curvature of the three-dimensional
spatial geometry. However, the intrinsic curvature of the one-dimensional
space is zero and thus $C_2$ ``decouples'' from it.
We will leave it as a coupling constant that determines an integration
constant in $l(\phi)$.
In order to obtain an exact solution, one has to solve (54) and (55).
$\dot\psi(t)$ in (56) can be re-written as
$$
C_0\dot\psi=2{{\dot a}^2\over a}-2\ddot a-2b{\dot\phi}^2a, \eqno(57)
$$
which is not necessarily integrable. None of the equations
of motion and other physical quantities ({\sl{e.g.}},
energy density and pressure in (59) and (60)) implicitly
or explicitly depend on $\psi(t)$. It is sufficient to
calculate $\dot\psi(t)$ in the following cases.

Now we see that when a desired $a(t)$ and an invertible $\phi(t)$
are given, they together yield $V(\phi)$ and $l(\phi)$ in (54) and (55).
We emphasize that especially in $4D$ single scalar cosmologies, this kind of
approach
(finding the matter to give a desired geometric state) has attracted
some attention (see [19,37] for discussions) due to the
fact that the metric field has only one degree of freedom (the scale
factor) and it depends on one variable $t$ only, making the field equations
relatively easy to handle. However with just one scalar this
approach (in two or four dimensions) yields exact $V(\phi)$ and $\phi(t)$
only for certain choices of the scale factors. In two
dimensions, we see from (54) and (55) that when $\psi$ and $l(\phi)$ are
present,
any choice of $a(t)$ and invertible $\phi(t)$ is possible.

For non-static solutions, the definition of mass (a conserved charge) is
no longer meaningful. Rather we treat the cosmological fluid as a perfect
fluid -- this can always be done in an $FRW$ model. When putting $T_{\mu\nu}$
in (8) in the form
$$
T_{\alpha\beta}=(p+\mu)U_{\alpha}U_{\beta}+pg_{\alpha\beta}, \eqno(58)
$$
where $U^{\alpha}$ is the normalized ($U^{\alpha}U_{\alpha}=-1$) two-velocity,
the energy density ($\mu$) and pressure ($p$) are given by
$$
\mu=-b{\dot\phi}^2-{1\over 2}V-{1\over 2}l{\dot\psi}^2, \eqno(59)
$$
and
$$
p=-b{\dot\phi}^2+{1\over 2}V-{1\over 2}l{\dot\psi}^2. \eqno(60)
$$
{}From (54), (55) and (56), (59) becomes
$$
\mu=-\bigl({\dot a\over a}\bigr)^2+{C_2\over 2a^2}, \eqno(61)
$$
and the pressure (60) can be written as
$$
\mu-p=-2{\ddot a\over a}. \eqno(62)
$$
One sees that if $C_2\leq 0$, then $\mu\leq 0$.
The weak energy condition ($WEC$) is always violated.
We will later see that in $4D$ $FRW$ cosmology, $\mu(t)$
always respects the $WEC$ even if the analogous $C_2$ vanishes.
We express the pressure in terms of $\mu-p$,
which plays the role as a gravitational mass.
When the strong energy condition $SEC$ (positivity of gravitational mass)
is respected, we see that $\ddot{a}<0$.

We illustrate three interesting examples. First of all, we consider
a non-singular scale factor
$$
a=\Bigl(1+{t^2\over A^2}\Bigr)^{1\over 2}, \qquad A^2>0. \eqno(63)
$$
This scale factor is an exact solution in a string-motivated single
scalar-field with a two-term exponential potential cosmology
in four dimensions [38]. Instead of adopting the fact that
$\phi\propto ln\bigl(1+{t^2\over A^2}\bigr)$ obtained in [38],
we first consider $\phi=kt$ for simplicity and illustrating the
freedom of choosing any invertible $\phi(t)$ for a given $a(t)$.
Now (54) is satisfied if
$$
V={2k^2\over (1+\phi^2)^2}, \eqno(64)
$$
where we have set $A^2k^2=1$. From (55) $l(\phi)$ is
$$
{C_0^2\over l}=2{k}^2
\Bigl({-1+\phi^2-b(1+\phi^2)^2\over 1+\phi^2}\Bigr)-C_2. \eqno(65)
$$
We can rewrite $a(t)$ in the form
$$
a(t)=(1+\phi^2)^{1\over 2} \quad .  \eqno(66)
$$
When $k=C_2=0$, $V(\phi)$, $\phi(t)$, $\mu(t)$ and $p(t)$ (see (67))
vanish; $a_o$ also vanishes and implying $\psi(t)$ is a constant.
Thus once $\phi(t)$ is ``switched off'',
the spacetime becomes a $2D$ Minskowski space. When $\psi(t)$ is switched off,
one must have $C_0=0$. Then (65) implies either $k=C_2=0$
or $\phi$ is a contant satisfying
$2k^2(-1+\phi^2-b(1+\phi^2)^2)=C_2(1+\phi^2)$.
In either case, the metric becomes Minskowski.
Therefore switching $\phi(t)$ or $\psi(t)$ off may lead to a Minskowski
spacetime.
Using (61) and (62), the energy density and pressure are easily calculated to
be
$$
\mu={k^2t^2(-2k^2+C_2)+C_2\over (1+k^2t^2)^2},
\qquad \mu-p=-{2k^2\over (1+k^2t^2)^2}. \eqno(67)
$$
$\mu(t)$ and $p(t)$ are similar to those obtained in [38] in terms of $t$
since they can be calculated from the scale factor directly
in two and four dimensions (see (85) and (86) later).
We will assume that $C_2>2k^2$ since the $WEC$ is respected for this
inequality.
The $SEC$ is always violated and therefore one gets an anti-gravitational
effect. It is clear that $\mu$ and $p$ are finite for all $-\infty<t<\infty$.
As $t\rightarrow\pm\infty$, $p=\mu\rightarrow {-2k^2+C_2\over k^2t^2}$
which is the equation of state of radiation in two dimensions [7,10].
At $t=0$, $\mu=C_2$ and $p=2k^2+C_2$. There is no
initial curvature and density singularity.

Now if we choose $\phi=\epsilon_1 ln\bigl(1+{t^2\over A^2}\bigr)$ as in [38],
$\mu(t)$ and $p(t)$ will be the same as before but with $k^2\rightarrow A^{-2}$
in (67). $V(\phi)$ now becomes
$$
V={2\over A^2}e^{-{2\over\epsilon_1}\phi}. \eqno(68)
$$
Comparing with the $4D$ case [38], $V(\phi)$ no longer depends two
exponential terms but just one exponential term. $l(\phi)$ is given by
$$
{C_0^2\over l}={2\over A^2}(1-4b\epsilon_1^2)-{4\over
A^2}(1-2b\epsilon_1^2)e^{-{\phi\over\epsilon_1}}-C_2. \eqno(69)
$$
Therefore, this $2D$ non-singular model has a similar scale factor,
energy density and pressure in terms of $t$ to that of
the $4D$ non-singular model. However, it
has just a single exponential potential term rather than two,
and an additional spectator field with the non-trivial coupling function.
In (63), when $A^{-2}\neq 0$, the birth of the universe
may be viewed as a quantum tunneling effect from $a=0$ (at an imaginary time),
to a bounce point $a=1$ (at $t=0$ where $\mu=C_2$ and $p=2A^{-2}+C_2$),
beyond which the universe evolves according to (63).

Note that without $l(\phi)$ and $\psi(t)$, one has to solve for $\phi(t)$
({\sl {i.e.}}, given a desired $a(t)$ in (55), we no longer have
the freedom to set $\phi(t)$ to whatever we want to simplify
the calculations -- it must be solved for). Simple calculations
show that $\dot\phi(t)$ can exactly be integrated in (55) for
$a(t)$ in (63), but the resultant $\phi(t)$ is not invertible.

Another case of non-singular universe model is described by the
following scale factor obtained in [38]:
$$
a=1+{t^2\over 2A^2}, \qquad A^2>0. \eqno(70)
$$
Again we assume that $\phi=kt$. Now (54) yields
$$
V={4k^2\over 2+\phi^2}, \eqno(71)
$$
where we have set $A^2k^2=1$. Now the coupling function $l(\phi)$ in (55)
is given by
$$
{C_0^2\over l}=k^2\Bigl((1-2b)\phi^2-2(1+b)-{b\phi^4\over 2}\Bigr)-C_2.
\eqno(72)
$$
The energy density and pressure in (61) and (62) become
$$
\mu={2C_2-4k^4t^2\over (2+k^2t^2)^2}, \qquad \mu-p=-{4k^2\over (2+k^2t^2)^2}.
\eqno(73)
$$
This model is less interesting than the previous one since the
$WEC$ is only respected for the range
${\sqrt{C_2\over 2}}{1\over k^2}>t>-{\sqrt{C_2\over 2}}{1\over k^2}$.
Note that the $WEC$ is respected in both models for all $t$ in four dimensions
[38].
$\mu(t)$ and $p(t)$ are finite for all $t$. As $t\rightarrow\pm\infty$,
$p=\mu\rightarrow -{4\over t^2}$. When $t=0$, $\mu={C_2\over 2}$
and $p=k^2+{C_2\over 2}$. The universe is
non-singular in every physical and geometical property and the
$SEC$ is violated. The universe may quantum mechanically evolve
from $a(t)=0$ to $a(t)=1$. All properties of this universe are
similar to the previous one except the $WEC$ is only respected
at a certain range of $t$.

Now if we choose $\phi=\epsilon_2 ln\bigl(1+{t^2\over 2A^2}\bigr)$
as in the case in [37], we get
$$
V={2\over A^2}e^{-{\phi\over\epsilon_2}}, \eqno(74)
$$
and
$$
{C_0^2\over l}={2\over A^2}(1-2b\epsilon_2^2)e^{\phi\over\epsilon_2}
+{4\over A^2}(b\epsilon_2^2-1)-C_2. \eqno(75)
$$
Thus the potential depends only on one exponential term.
The simplier $V(\phi)$ in (68) and (74) compared with the two-term
potential in four dimensions in [38] is due to the presence of $l(\phi)$
in (65) and (72) and the simplier field equation (54) in two dimensions.
If $l(\phi)$ and $\psi(t)$ are constant,
it can be checked that $\dot\phi(t)$ can still be integrable in (55) for
$a(t)$ given in (70). However, the resultant $\phi(t)$ is not invertible.
We note that a previous attempt on construction of singularity-free $2D$
cosmologies was considered in [38], where the authors used the
same kind of action (but a different potential)
they considered in [20] for a
construction of a non-singular black hole metric
(see the discussion in the introductory section in this paper).
In the cosmological case, they
obtained $FRW$ solutions that are non-singular and asymptotically
approach a dust-dominated universe at late time. However, explicit
solution for the scale factor is not possible. In our cases, all
$a(t)$, $\phi(t)$, $l(\phi)$ and $V(\phi)$ are exact (it can be checked
that $\psi(t)$ are exact in the above two cases as well).
No approximations and asymptotic limits are taken.

Our final example is a singular universe with some interesting properties.
The desired scale factor is given by
$$
a=ABt^{1\over 2}+At^{2\over 3}=At^{1\over 2}(B+t^{1\over 6}), \qquad A,B>0
\quad .\eqno(76)
$$
As $t^{1\over 6}\ll B$, $a(t)\rightarrow ABt^{1\over 2}$ which is the scale
factor
for a radiation-dominated universe in four dimensions. As $t^{1\over 6}\gg B$,
$a(t)\rightarrow At^{2\over 3}$ which is the scale factor of a dust-dominated
universe in four dimensions. Physically this means that for early enough times,
the $2D$ universe behaves as a $4D$ radiation universe and for sufficiently
late
times it behaves like a $4D$ dust universe. Thus we have a $2D$ cosmological
model which resembles the classical evolution (without any kind of inflation)
of a $4D$ radiation/dust universe. We need $\phi$ to complete our
solution. For simplicity, we set $\phi=ln\bigl({t\over t_o}\bigr)$ and
$-4b={41\over 18}$. Now , $V(\phi)$ in (54) is given by
$$
V(\phi)=-{{B\over 2}t_o^{1\over 2}e^{\phi\over 2}
        +{4\over 9}t_o^{2\over 3}e^{2\phi\over 3}\over
        Bt_o^{1\over 2}e^{\phi\over 2}+t_o^{2\over 3}e^{2\phi\over 3}}.
\eqno(77)
$$
$l(\phi)$ in (55) is given by
$$
{C_0^2\over l}=-{5A^2B^2\over 36t_o}e^{-\phi}+{7A^2\over 36t_o^{2\over
3}}e^{-{2\over 3}\phi}-C_2. \eqno(78)
$$
The energy density and pressure become
$$
\mu=-{(B+{4\over 3}t^{1\over 6})^2\over 4(B+t^{1\over 6})^2t^2}+{C_2\over
2A^2t(B+t^{1\over 6})^2}, \eqno(79)
$$
$$
\mu-p={B+{8\over 9}t^{1\over 6}\over 2(B+t^{1\over 6})t^2}. \eqno(80)
$$
Both $\mu(t)$ and $p(t)$ diverge at $t=0$. The curvature scalar
diverges as well. When $B\gg t^{1\over 6}$,
$\mu\rightarrow -{1\over 4t^2}+{C_2\over 2A^2B^2t}$, $p=-{3\over 4t^2}
+{C_2\over 2A^2B^2t}$. When $B\ll t^{1\over 6}$,
$\mu\rightarrow -{4\over 9t^2}+{C_2\over 2A^2t^{4\over 3}}$
and $p=-{8\over 9t^2}+{C_2\over 2A^2t^{4\over 3}}$.
A graphical analysis of (79) shows that
the $\mu<0$ in the range $0<t<t_w$ (one sets $\mu=0$ in (79)
and the real positive root will be the $t_w$; however due to
the cubic power in $t$, one cannot generally get the expression $t_w$
explicitly in terms of $A$, $B$ and $C_2$) and $\mu>0$ for $t_w>0$.
Thus, for this ``big bang'' model the $WEC$
is respected only at a certain range of time.
The $SEC$ is not violated, making the active gravitational mass positive,
and therefore, the above ``$big-bang\rightarrow radiation{\rightarrow}dust$''
singular universe possible.

Motivated by the similarities in terms of field
equations between $2D$ and $4D$ $FRW$ cosmological
models, we briefly discuss a two-scalar $FRW$ model in four dimensions.
We expect that due to the more complicated nature of the field equations,
it is more difficult to get exact solutions even through we follow the
above approach to solve the field equations. However, we can still get
an exact dust/radiation solution. The $4D$ action we consider is
$$
S=\int d^4x{\sqrt{-g}}\bigl(R-{1\over
2}l(\phi)(\nabla\psi)^2-2(\nabla\phi)^2-2V(\phi,\psi)\bigr).
\eqno(81)
$$
We will consider the choice $l=e^{4\phi}$ in the above action where
it corresponds to the bosonic part of the low energy heterotic
string action to zero order in the string tension $\alpha^{'}$ [40].
To order $\alpha^{'}$, higher order curvature
terms such as Gauss-Bonnet curvature will be present. For simplicity,
we only consider zero order in $\alpha^{'}$ and ignore these terms.
Again $\psi(t)$ is the spectator and $\phi(t)$ is
the dilaton. The exact form of the dilaton-spectator potential
$V(\phi,\psi)$ is not known in string theory [40].
We will derive the form of it for a desired scale factor instead of
assuming it {\it a-priori}. In order to simplify the calculations,
we assume $V(\phi,\psi)=V(\phi)$.
The cosmological fluid ({\sl{e.g.}} dust and radiation) is described
by $\phi(t)$ and $\psi(t)$.

{}From (81) the field equations and conservation equations are
$$
3H^2+3{C_2\over a^2}={\dot\phi}^2+{1\over 4}e^{4\phi}{\dot\psi}^2+V. \eqno(82)
$$
$$
3{\ddot a\over a}=V-2{\dot\phi}^2-{1\over 2}e^{4\phi}{\dot\psi}^2, \eqno(83)
$$
where
$$
{\dot\psi}={2C_0\over a^3}e^{-4\phi}, \eqno(84a)
$$
is the spectator equation and the dilaton equation is
$$
\ddot\phi+3H\dot\phi+{1\over 2}{dV\over d\phi}-{1\over 8}{dl\over
d\phi}{\dot\psi}^2=0. \eqno(84b)
$$
The first one is the Friedmann equation. The second one is
Raychaudhuri's equation. $H={{\dot a}\over a}$ is the Hubble parameter.
The third one is the spectator equation.
The constant $C_2$, which has dimensions of inverse length squared
has the following significance.
For either a dust- or radiation-dominated universe,
when $C_2$ is positive, the universe is spatially closed and
it starts with a big bang and ends at
a big crunch, while a vanishing (or negative) $C_2$ implies a forever-
expanding spatially open universe. In two dimensions, however,
$C_2$ has no relationship with the one-dimensional spatial geometry.
It only relates to the positivity of the energy density in (61).
Provided that ${\dot\phi}\neq 0$, the last equation
follows from the conservation of stress-energy,
$\nabla^{\mu}T_{\mu\nu}=0$, where $T_{\mu\nu}$
is in the perfect fluid form with energy density and pressure given by
$$
\mu={\dot\phi}^2+{1\over 4}e^{4\phi}{\dot\psi}^2+V=3{C_2\over a^2}
+3\bigl({\dot a\over a}\bigr)^2. \eqno(85)
$$
$$
p={\dot\phi}^2+{1\over 4}e^{4\phi}{\dot\psi}^2-V
=-\Bigl({C_2\over a^2}+2{\ddot a\over a}+\bigl({\dot a\over a}\bigr)^2\Bigr).
\eqno(86)
$$
Comparing (85) with (61), we see that it is ``easier'' for
the $2D$ models to violate the $WEC$ for a given scale factor and
$C_2$ (with arbitrary signs).
If we obtain a solution of (82), (83) and (84) with $\dot\phi\neq 0$,
the local consevation equation $\nabla_{\mu}T^{\mu\nu}=0$
will necessarily be satisfied. Again $\dot\psi$ in (84) is not required to be
integrable, since we have assumed $V(\psi,\phi)=V(\phi)$. To
specify a model, we note that twice (82) plus (83) gives
$$
V=2H^2+{2C_2\over a^2}+{\ddot a\over a} \eqno(87)
$$
while the combination (82)-(83) gives
$$
{\dot\phi}^2=-{C_0^2\over a^6}e^{-4\phi}+{C_2\over a^2}-{\dot H}. \eqno(88)
$$
These two equations together are equivalent to the system (82)--(84).
It is generally not trivial to get exact solutions for (87) and (88).
It is particularly interesting to see whether (87) and (88) can yield
dust/radiation scale factors. For this purpose we try the ansatz
$$
a=At^n. \eqno(89)
$$
Now, $H={n\over t}$, ${\dot H}=-{n\over t^2}$ and
${{\ddot a}\over a}={n(n-1)\over t}$. In order to
solve (87) and (88), we choose $\phi={\alpha}ln\bigl({t\over\beta}\bigr)$.
An exact solution can be obtained when $C_2=0$ and
$$
(-1+n)(1-9n)={4C_0^2\beta^{4\alpha}\over A^6}, \qquad 2\alpha=1-3n \quad .
\eqno(90)
$$
{}From this equation, we see that we must have $1>n>{1\over 9}$. The potential
in (87) is given by
$$
V={n(3n-1)\over\beta^2}e^{-{2\over\alpha}\phi}. \eqno(91)
$$
(85) and (86) imply that $\mu={3n^2\over t^2}$ and $p={n(2-3n)\over t^2}$.
When $n={1\over 2}$ and ${2\over 3}$, one obtains radiation and
dust scale factor respectively. The potentials are ${1\over 4\beta^2}e^{8\phi}$
for the radiation and ${2\over 3\beta^2}e^{4\phi}$ for dust.
Earlier attempts on modelling radiation/dust universe
using a single scalar can be found in [19].
Here we obtain the same result but the differences are
that the action (81) is inspired from low energy string theory (except that
the potential is arbitrary) and we have two scalars instead of one.

Finally we comment that if we assume a general coupling function
$l(\phi)$ instead of $e^{4\phi}$ in action (81) ({\sl{i.e.}},
we no longer consider string theory), then using
(87) and (88) it is easy to see that we can have any desired $a(t)$ and
invertible
$\phi(t)$ by adjusting $l(\phi)$ and $V(\phi)$, similar to the $2D$ solutions
we discussed above. More precisely, given any $a(t)$ and an invertible
$\phi(t)$, one uses (87) and (88) to trivially obtain $V(\phi)$ and $l(\phi)$
respectively. Therefore this $4D$ model is very useful for
constructing the cosmological fluid required to obtain desired features of
a $4D$ universe. For example, it is interesting to
implement the big bang model $a(t)$ in (76) in the $4D$ action (81).
Further discussion on this point is beyond the
scope of this paper. In [19], where $C_0=0$ in (88),
the authors could only solve for $\phi(t)$ for limited choices of $a(t)$ in
(88).
The reason is that $\phi(t)$ may not be solvable, or even if
solvable, is non-invertible for a desired $a(t)$ when $C_0=0$ in (88).
Neverthless it would be of interest and to consider a two- or even multiple-
scalar system as a cosmological fluid and then
investigate under what conditions the
desired features of our universe can be reproduced.

\vskip 0.6 true cm

\noindent{\bf{6 Conclusions}}

\vskip 0.3 true cm

The common means of determining the classical metric of a given spacetime is to
begin with a prescribed action (that is typically motivated by some set of
fundamental
physical principles and symmetry requirements) and then find exact solutions to
its associated
field equations for boundary conditions of physical interest.

In this paper we have taken what is essentially the inverse approach: within
the context of the
$(1+1)$ dimensional $R=T$ theory we have shown that it is possible to find the
matter required to give a desired geometry of spacetime.  The matter contains
the dilaton $\phi$
and spectator $\psi$ fields with a coupling function $l(\phi)$
and a dilaton potential $V(\phi)$ in (2). In the usual approach,
treating $\phi$ and $\psi$ as the matter source does not
determine the dynamics until $V(\phi)$ and $l(\phi)$ are known;
in our approach we take the advantage of this arbitrariness to determine the
$V(\phi)$ and $l(\phi)$ necessary to generate our desired solution.
In essence we are
`engineering'  the black hole solutions we want,
as opposed to `discovering' them as solutions
to an {\it a-priori} action. Although this procedure does not
produce generic restrictions on $V(\phi)$ and $l(\phi)$,
it does necessitate that the
masses of the desired black holes are determined in terms of the coupling
constants
in the potentials, as opposed to being fully arbitrary.
Such black holes are similar to discrete bound state solutions, in contrast
with black holes in ordinary general relativity.
The $V(\phi)$ and $l(\phi)$ so
determined may therefore have other interesting solutions -- we have not
searched
for these in this paper.

The black hole metrics we have considered in this paper are of
physical interest insofar as they model interesting spacetime
structures of physical relevance in higher dimensions or spacetimes
that arise from low-energy effective string-inspired actions.  Specifically,
we have been able to construct exact black hole solutions with
double and multiple horizons and have computed their Hawking temperature and
entropy. One of the solutions resembles
the $2D$ string black hole with the same thermodynamics.
A $2D$ version of the Reissner-Nordstrom solution was also
constructed. However, there are two features of all
the black hole solutions. As mentioned above, the solution must be such that at
least some of the coupling constants in $l(\phi)$ (or $V(\phi)$ in some
cases) functionally determine an integration constant (namely the mass
parameter).
Second, when a test particle is linearly
coupled to $\psi$ and $\psi^{'}$, it will respectively encounter
an infinite tidal force at the horizon or an infinite potential
barrier outside the horizon. In the $2D$ string black holes,
no such infinities are present.

In the context of $FRW$ cosmology, we represented the
cosmological fluid by $\phi$ and $\psi$ again.
Given any desired $a(t)$ and invertible
$\phi$, the field equations are exactly solvable
by adjusting $V(\phi)$ and $l(\phi)$. We illustrated
two non-singular solutions which respectively resemble
two $4D$ string-motivated cosmological solutions. A
singular model which is analogous to $4D$ standard
big-bang/radiation/dust was also constructed. In addition,
we solved the equations of motion for a $4D$ low energy
string action (81) and obtained a dust and a radiation
solutions. We note that if the coupling function in (81) is left arbitrary
instead of $-{1\over 2}e^{4\phi}$, one can have any desired $a(t)$ as an
exact cosmological solution by adjusting the dilaton potential
and the coupling functions, similar to the two-dimensional cases.

Finally we comment on two facts. First of all, due to the simplicity
of gravitational field equations in two dimensions,
we see that one can have two different $2D$ gravitational theories (
$R=T$ and string) that both yield the same classical black hole geometry
and thermodynamical properties.
It is tempting to see whether one can
construct a new $2D$ gravitational theory which can reproduce the various
$2D$ string black hole solutions but is free
of dilaton divergences and setting integration constants equal to  coupling
constants.
The new theory may serve as another toy model for investigation of
black hole information loss.
Second, given an asymptotically flat black hole solution which has
double or multiple horizons, it is natural to ask whether or not exact mass
inflation can happen. Mass inflation is a well-known phenomenon in four
dimensions. More recently, mass inflation has been shown to occur in
$2D$ and $3D$ string black hole solutions (see {\sl{e.g}} [41]).
For the black hole solutions in this paper, which share the same classical
geometry and thermodynamics properties with the string cases,
it is very difficult, if not impossible,
to solve the non-static field equations
in the Eddington-Finkelstein ingoing co-ordinates [42]. Even
if the field equations could be solved exactly, it is difficult to
interpret the mass-inflating parameter in action (2).
It was claimed in [43] that mass inflation is a generic phenomenon,
and does not depend on the particular details of the theory. It would
be interesting
to see whether or not the double-horizon black hole solutions
in this paper may serve as counter-examples.
We intend to relate further details elsewhere.

\vskip 0.6 true cm

\noindent{\bf{Acknowledgement}}

\vskip 0.3 true cm

This work was supported by the Natural Science and Engineering
Research Council of Canada. We would like to thanks
Jimmy Chan for writing a MAPLE computer program to
calculate the constant $C$ in (15) and discussing the difficulties of
mass inflation of the black hole solutions in this paper.
K.C.K. Chan also wishes to thanks Amy Leung for checking the manuscript.
R.B. Mann would like to thank the hospitality of D.A.M.T.P. in
Cambridge, where some of this work was carried out.

\vskip 0.6 true cm

\noindent{\bf{References}}

\vskip 0.3 true cm

\item{1.} Mann R B 1992 {\sl{Gen. Rel. Grav.}} {\bf{24}} 433
\item{2.} Mann R B 1991 {\sl{Found. Phys. Lett.}} {\bf{4}} 425
\item{3.} Mann R B, Morsink S M, Sikkema A E and Steele T G 1991
{\sl{Phys. Rev.}} D {\bf{43}} 3948; Mann R B 1992 {\sl{Phys. Lett.}} B
{\bf{294}} 310
\item{4.} Witten E 1991 {\sl{Phys. Rev.}} D {\bf{44}} 314;
Mandal G, Sengupta A and Wadia S 1991 {\sl{Mod. Phys. Lett}} A {\bf{6}} 1685
\item{5.} Mann R B and Ross S F 1993 {\sl{Class. Quantum Grav.}} {\bf{10}} 1405
\item{6.} Lemos J P S and Sa P M 1994 {\sl{Class. Quantum Grav.}} {\bf{11}} L11
\item{7.} Sikkema A E and Mann R B 1991 {\sl{Class. Quantum Grav.}} {\bf{8}}
219
\item{8.} Morsink S M and Mann R B 1991 {\sl{Class. Quantum Grav.}} {\bf{8}}
2257
\item{9.} Chimento L P and Cossarini A E 1993 {\sl{Class. Quantum Grav.}}
{\bf{10}} 2001
\item{10.} Chan K C K and Mann R B 1993 {\sl{Class. Quantum Grav.}} {\bf{10}}
913
\item{11.} Jackiw R 1984 {\sl{Quantum Theory of Gravity}} ed S Christensen
(Bristol: Hilger) p 403
\item{12.} Achucarro A and Ortiz M E 1993 {\sl{Phys. Rev.}} D {\bf{48}} 3600
\item{13.} Lechtenfeld O and Nappi C 1992 {\sl{Phys. Lett.}} B {\bf{288}} 72
\item{14.} Elizalde E and Odintsov S D 1993 {\sl{Nucl. Phys.}} B {\bf{399}} 581
\item{15.} Cadoni M and Mignemi S 1994 {\sl{Nucl. Phys.}} B {\bf{427}} 669
\item{16.} Mann R B 1994 {\sl{Nucl. Phys.}} B {\bf{418}} 231
\item{17.} Cadoni M and Mignemi S 1993 {\sl{Phys. Rev.}} D {\bf{48}} 5536;
Antoniadis I, Rizos J and Tamvakis K 1994 {\sl{Nucl. Phys.}} B {\bf{415}} 497;
Cvetic M and Tseytlin A A 1994 {\sl{Nucl.Phys.}} B {\bf{416}} 137
\item{18.} Berkin A L and Hellings R W 1994 {\sl{Phys. Rev.}} D {\bf{49}} 6442
\item{19.} Ellis G F R and Madsen M S 1991 {\sl{Class. Quantum Grav.}} {\bf{8}}
667
\item{20.} Trodden M, Mukhanov V and Brandenberger R 1993 {\sl{Phys. Lett.}} B
{\bf{316}} 483
\item{21.} Izawa K-I 1994 {\sl{Prog. Theo. Phys.}} {\bf{91}} 393
\item{22.} Romans L J, 1992 {\sl{Nucl. Phys.}} {\bf B383} 395
\item{23.} Kalligas D, Wesson P, and Everitt C W F 1992 {\sl{Gen. Rel. Grav.}}
{\bf{24}} 351
\item{24.} McGuigan M D, Nappi C R and Yost S A 1992 {\sl{Nucl. Phys.}} B
{\bf{375}} 421
\item{25.} Mann R B Shiekh A and Tarasov L 1990 {\sl{Nucl. Phys.}} B {\bf{341}}
134
\item{26.} Mann R B 1993 {\sl{Phys. Rev.}} D {\bf{47}} 4438
\item{27.} Brown J D and York J W 1993 {\sl{Phys. Rev.}} D {\bf{47}} 1407;
  Brown J D, Creighton J and Mann R B 1994 {\sl{Phys. Rev.}} D {\bf{50}} 6349
\item{28.} Criegton J (unpublished)
\item{29.} Nappi C R and Pasquinucci A 1992 gr-qc/9208002 (unpublished)
\item{30.} Perry M J and Teo E 1993 {\sl{Phys. Rev. Lett.}} {\bf{70}} 2669.
\item{31.} Mann R B, Morris M S and Ross S F 1993 {\sl{Class. Quantum Grav.}}
{\bf{10}} 1477
\item{32.} Bars I and Sfetsos K 1992 {\sl{Phys. Rev.}} D {\bf{46}} 4510
\item{33.} Bekenstein J D 1975 {\sl{Ann. Phys. (N.Y.)}} {\bf{91}} 75
\item{34.} Wald R  1993 {\sl{Phys. Rev.}} D {\bf{48}} 3427
\item{35.} Chimento L P and Cossarini A E 1994 {\sl{Class. Quantum Grav.}}
{\bf{11}} 1177
\item{36.} Bento M C and Bertolami O 1994 {\sl{Phys. Lett}} A {\bf{193}} 31
\item{37.} Bayin S S, Cooperstock F I and Faraoni V 1994 {\sl{Astrophys. J.}}
{\bf{428}} 439
\item{38.} Ozer M and Taha M O 1992 {\sl{Phys. Rev.}} D {\bf{45}} R997
\item{39.} Moessner R and Trodden M 1994 gr-qc/9405004
\item{40.} O'Neill C M 1994 {\sl{Phys. Rev.}} D {\bf{50}} 865
\item{41.} Chan J S F and Mann R B 1994 {\sl{Phys. Rev.}} D {\bf{50}} 7376;
          Chan J S F, Chan K C K and Mann R B gr-qc/9406049
\item{42.} Chan J S F (private communications)
\item{43.} Droz S 1994 {\sl{Phys. Lett.}} A {\bf{191}} 211
\bye